\documentclass[11pt]{article}

\usepackage[T1]{fontenc}
\usepackage[margin=1in]{geometry}
\usepackage{setspace}
 \usepackage{authblk}

\usepackage{amsmath,amssymb,amsfonts}
\usepackage{algorithmic}
\usepackage{graphicx}
\usepackage{textcomp}
\usepackage{mathrsfs}
\usepackage{mathtools}
\usepackage{dsfont}
\usepackage{caption}
\usepackage{subcaption}
\usepackage{hyperref}
\usepackage{paralist}
\usepackage{bbold} 
\usepackage{pgfplots}
\pgfplotsset{compat=1.18}

\usepackage{comment} 
\usepackage{pgf}
\usepackage{import}
\usepackage{tikz}

\usepackage{multirow}

\newcommand{\proofof}[1]{\textit{\textbf{Proof of #1:}} }

\def\QEDclosed{\mbox{\rule[0pt]{1.3ex}{1.3ex}}} % for a filled box
\def\QED{\QEDclosed} % default to closed

\def\BibTeX{{\rm B\kern-.05em{\sc i\kern-.025em b}\kern-.08em
    T\kern-.1667em\lower.7ex\hbox{E}\kern-.125emX}}

\title{\LARGE \bf Local Reproduction Numbers for Analysis and Optimal Mitigation of Epidemic Processes}
\author[1]{Jos\'{e} I. Caiza\thanks{Corresponding author: José I. Caiza (jcaiza@purdue.edu), Junjie Qin (jq@purdue.edu), Philip E. Paré (philpare@purdue.edu). This work was supported in part by the National Science Foundation, grants
NSF-ECCS \#2032258, \#2238388, and \#2339803.}}
\author[1]{Junjie Qin}
\author[1]{Philip E. Par\'{e}}
\affil[1]{Elmore Family School of Electrical and Computer Engineering, Purdue University}
\date{}

\newcommand{\jose}[1]{\textcolor{black}{#1}}
\newcommand{\diag}{\text{diag}}

\newcommand{\pop}{\mathcal{P}}
\newcommand{\infra}{\mathcal{I}}
\newcommand{\intra}{\text{Intra}}

\newcommand{\HR}{\text{H}}
\newcommand{\TR}{\text{T}}

\newtheorem{corollary}{\textbf{Corollary}} 
\newtheorem{assumption}{ \textbf{Assumption}}
\newtheorem{lemma}{\textbf{Lemma}}
\newtheorem{theorem}{\textbf{Theorem}}
\newtheorem{proposition}{\textbf{Proposition}}
\newtheorem{problem}{\textbf{Problem}}
\newtheorem{definition}{\textbf{Definition}}

\newtheorem{remark}{\textbf{Remark}}

\begin{document}
\maketitle

\begin{abstract}
In this paper, we propose a unifying framework to analyze layered networked epidemic processes where the population, represented either as a networked SIS or SIR model, is coupled with an infrastructure network. We propose novel local metrics that enable the analysis of the epidemic spreading locally (node-level) and the prediction of the %network-level metrics, namely, basic and effective reproduction number,
network-wide spreading under some sufficient conditions. Inspired by the bilayer nature of our model, we partition the layered network between intra- and inter-layer interactions. We identify conditions at the layer level which allow the %closed-form solution  
characterization of the network-level basic reproduction number %bridging the gap between 
in terms of the physical structure of the layered network%and its epidemic spreading 
. Then, we leverage the local metrics to propose an optimization framework for epidemic mitigation based on a fundamental notion: the network-wide spreading can be quantified by the aggregation of the local (node-level) epidemic burden. We provide guarantees that the local metric-based optimization can be solved as a geometric program. We illustrate our theoretical results via numerical examples and evaluate the performance of the local metrics and the optimization framework they enable versus other existing epidemic mitigation approaches based on network-level metrics.

\vspace{0.5em}
\end{abstract}

\section{Introduction}
Epidemics can be catastrophic, causing worldwide impacts from human losses to financial crises~\cite{WHO}. The Covid-19 pandemic disclosed the importance of developing tools to accurately capture the behavior of disease-spreading event, and as a byproduct, possible optimal mitigation decisions~\cite{harvard,policyCOVID19}. Developing epidemiology models and efficient control policies have been questions of longstanding interest in the scientific community as they play a pivotal role in assisting public health officials in their policy-making against epidemic outbreaks. 
%\subsection{Related Work}
%\subsection{Proposed Work and Its Contributions}
%\subsection{Related Work}

Networked SIS and SIR models have been studied extensively in the controls community in recent years \cite{hethcote1989three,pare2020analysis,MEI2017116}. However, a vast majority of traditional networked models only account for person-to-person interaction as the propagation mechanism. Nevertheless, diseases can also propagate through other means, such as water distribution systems \cite{RANDAZZO2020115942,VERMEULEN2015109} or contaminated environments (e.g. public transportation services \cite{brooks} and hospitals \cite{CICCOLINI2013380}). Consequently, new models have been recently proposed where a water compartment is coupled with traditional, single-population epidemic compartmental models~\cite{Tien2010-ch, SHUAI2011118}. %, a Susceptible-Infected-Water-Removed (SIWR) compartmental model was proposed by coupling the classical SIR person-to-person infection with a contaminated water compartment. In \cite{SHUAI2011118}, a cholera model was proposed where the virus propagates through both direct and indirect transmission pathways with a water compartment. 
In \cite{liu}, %the model proposed in \cite{Tien2010-ch} is extended by considering 
a networked SIS model coupled with a contaminated single water compartment (also known as resource node for generalization) was proposed and it was denoted as the \textit{SIWS model}. Later, the networked 
% formulation of the model in~\cite{liu} 
model
was developed in~\cite{pare2023multilayer} (layered networked SIWS model) and we introduced the layered networked SIWR model in~\cite{caiza2024}.

The analysis of the aforementioned systems depends widely on the use of network-level metrics, namely, the \textit{basic reproduction number} $R^0$ or the \textit{effective reproduction number} $R(t)$ (for a survey, see\cite{MEI2017116,pare2023multilayer,PARE2020345}). %More precisely, Paré et al. \cite{pare2023multilayer} provided $R_0$-based sufficient conditions for the global asymptotic stability of the healthy state and the endemic equilibrium. %However, the analysis based on $R_0$ does not enable the characterization of the dynamical behavior of the layered networked model. %Since we proved that the layered  networked SIWR model always attains the healthy equilibria, %In \cite{caiza2024}, we proved that the healthy equilibria is always attained for the proposed layered networked SIWR model and 
%Thus, a metric of interest in the literature is the \textit{network-level effective reproduction number $R^t$}, which, in contrast to $R_0$, is time dependent and includes the evolution of the susceptible states in a networked population
By definition, %$R^0$ does not account for the evolution of the susceptible proportion in group of individuals. Likewise, 
$R^0$ assumes that everyone is susceptible to the disease, whereas in real life, some people may already be protected or recovered from past infections. Unlike $R^0$, the \textit{effective reproduction number $R(t)$} %is a network-level metric that 
considers the evolution of the susceptible state, revealing how a disease spreads across a network in real time~\cite{MEI2017116,baike_peak_infection,caiza2024}. %we introduced in~\cite{caiza2024} the \textit{network-level effective reproduction number} $R^t$. 
% In %particular
% \cite{caiza2024}, we provided $R^t$-based sufficient conditions to predict the transient behavior of the %weighted average of the infection state of the layered networked SIWR model
% epidemic spreading in the layered networked SIWR model. 
While $R^0$ and $R(t)$ are efficient to analyze epidemic processes at the \textit{network-level}, they fail to describe the transient behavior locally (\textit{node-level}). This limitation is addressed by the novel \textit{local effective reproduction numbers (LERNs)} introduced in~\cite{baike2023DRNs,caiza2024}. %for networked SIS and SIR models. Building on~\cite{baike2023DRNs}, we introduced the \textit{local effective reproduction numbers} (LERNs)~\cite{caiza2024}, %for the layered networked SIWR model
 %which %enable the prediction of epidemic spreading at the node level and, 
 Under sufficient conditions, the LERNs 
 can be leveraged to predict the network-wide transient behavior of networked epidemic processes.
 % with recovery rates.
   %quantifies the aggregated expected number of infections resulting from the pair-wise interactions of a given node with all of its possible source of infection. %In this article, we aim to present a \textit{unifying framework} for the modeling and analysis of the \textit{layered networked SIWS and SIWR models}, both at the node and local-level. 

%\jose{In this paper, we provide a common structure that builds off our work in~\cite{caiza2024} and enables the analysis of the transient behavior at the network and node level of layered networked models where the population can be either modeled as a  networked SIS or SIR model (in our previous work~\cite{caiza2024} we only focused on SIWR). In particular, we delve into the particularities to predict the transient behavior of layered networked SIWR and SIWS based on the network-level ERN. Moreover, we extend the concept of LERNs proposed in~\cite{caiza2024} to define local metrics that enable the prediction of the network-level BRN.}

While the discussion insofar has been centered around modeling and analysis, another critical challenge that policy makers face is to optimally mitigate the epidemic spreading under physical constraints, i.e., maximum improvement in the mean recovery time  or transmission rates. %In this context, it is of particular interest to provide insights on the cost-optimal mitigation strategies given available control actions that provide protection to the layered networked model (e.g. increasing the healing rate or decreasing the infection rate). 
The strategies for epidemic mitigation have been addressed in the literature %for standard compartmental and networked epidemic models
focusing on three approaches. First, in the \textit{optimal control framework}, the minimization of a cost function integrated along a trajectory is attained by computing optimal parameters or control inputs~\cite{rowthorn, LEE2010136, HAYHOE2021495, somers, ACEMOGLU2024111391, Cianfanelli}. %using Pontryagin's minimum principle \cite{rowthorn, LEE2010136, HAYHOE2021495}. 
% The aforementioned approach generally requires numerical solutions of Pontryagin's optimality conditions \cite{rowthorn, LEE2010136}, large-scale receding horizon optimization to account for discrete dynamics \cite{HAYHOE2021495}, or linearization of a discounted cost function about the initial infection state~\cite{somers}. %In \cite{ACEMOGLU2024111391}, an optimal adaptive testing policy is characterized for a compartmental SIR model that includes testing as control action.  %In \cite{somers}, a convex optimization is proposed for the resource allocation in a networked SIS model where the discounted cost function is converted into a linear function of the initial condition of the infection state.
The second major approach is the \textit{spectral abscissa-based mitigation}, where the characterization of the Perron-Frobenius eigenvalue as a geometric program is leveraged to formulate constrained resource allocation problems. More precisely, the optimal allocation aims to reduce the spectral abscissa of the model's Jacobian matrix about some healthy equilibrium %subject to rate or budget constraints 
\cite{preciado2014, preciado2,mai,enyioha,ashish}. %The main advantage of this framework is that the computation of the trajectories are not needed to solve the optimal control problem. 
Lastly, the third approach involves the \textit{convex optimization of $R^0$}. In \cite{smith}, a geometric program characterization of $R_0$ is provided, enabling the formulation of $R_0$-based optimal resource allocation problems %under rate and budget constraints 
as constrained geometric programs (GPs). The main advantage of the spectral abscissa and $R_0$ optimization is that the computation of the trajectories are not needed to solve an optimal control problem. %Even though both the spectral abscissa and $R_0$ are equivalent threshold parameters to predict an epidemic spreading, they lead to different optimal allocations. %different quantities leading to different allocations, they are equivalent threshold parameters to predict an epidemic spreading. %The authors show through numerical examples the qualitative differences between optimizing $R_0$ instead of the spectral abscissa. 
In this paper, we extend our previous work on the analysis of layered epidemic processes \cite{caiza2024} by providing a more comprehensive analysis %of the transient behavior 
for both the layered networked SIWS and SIWR models at the node-level. %Moreover, we extend the novel concept of LERNs to define local metrics, which under sufficient conditions could be used to predict $R_0$. %Departing from the preceding layered networked epidemic models, we propose a \textit{unifying framework} to analyze the \textit{layered networked SIWS} and \textit{SIWR models}, both at the network and local (node) level.  
This framework is mainly motivated by the fact that the dynamical behavior of the layered infection state (i.e., infection proportion in group of individuals  %in the population network 
and contamination of resource nodes) %in the infrastructure network) 
in both layered networked models can be defined in terms of its susceptible state. %Whereas the aforementioned approaches are centralized, 
% Furthermore, we %seek to 
% leverage the novel local metrics we propose to formulate a distributed optimization framework for epidemic mitigation in the layered networked model. To the best of our knowledge, there is no work in the literature that has leveraged local metrics in allocation problems for epidemic mitigation.
Inspired by the physical notion used to define the LERNs, we propose novel time-independent local metrics that capture the epidemic burden at each node in the case of an epidemic outbreak, namely, the novel \textit{local basic reproduction numbers (LBRNs)}.  We aim to understand under which conditions the LBRNs can be leveraged to establish a bound for $R^0$ or compute its closed-form solution. Then, we employ the LBRNs to quantify the epidemic spreading at the network-level by  following a fundamental concept: the epidemic spreading across the layered network can be measured by the aggregated sum of the local epidemic burden. 

\textit{Article contributions:} The contributions of our paper is twofold: 1) the analysis of the epidemic spreading in the layered network based on local and layer-wise metrics and 2) the use of the LBRNs to formulate optimization problems for epidemic mitigation.
We summarize the contributions and main results of the proposed work as follows:
\begin{enumerate}
    \item First, %we provide a characterization of the weighted average of the layered infection state based on $R^t$ for both layered networked models. 
    we define the network-level $R^0$ and $R(t)$ for both layered networked SIWS and SIWR epidemic models. %in terms of its susceptible state
     Then, we identify LERNs-based conditions to predict the evolution of the epidemic spreading locally (node-level) and network-wide (see Theorems~\ref{theo:Rt_local}--\ref{theo:node_to_network}).
    \item We introduce the novel %\textit{local basic reproduction numbers (LBRNs)} 
    LBRNs to capture the epidemic risk locally. We leverage the LBRNs to determine a bound for $R^0$ and a sufficient condition %under heterogeneous parameters 
    to characterize $R^0$ %in terms of the LBRNs 
    under heterogeneous spreading (see Proposition~\ref{prop:gap_LBRNs}).
    \item Motivated by the bilayer nature of our model, we identify a condition to partition the layered network with respect to the intra- and inter-layer %: intra and inter-layer
    interactions. Thus, we characterize $R^0$ %of the layered network 
    by computing a 2-by-2 matrix, establishing a bridge between the physical structure of the layered network and its impact on $R^0$ (see Theorem~\ref{theo:layer_pair-wise}).
    \item We propose the novel LBRNs-based allocation framework for epidemic mitigation: budget and rate-constrained. %by using the aggregated %local epidemic burden (represented by 
    %sum of the LBRNs as the metric for the network-wide epidemic spread. 
    We provide guarantees that our framework can be effectively solved with off-the-shelf convex optimization techniques for GPs (see Theorem~\ref{theo:opt_budget}). 
\end{enumerate}  %Third, %exploiting the definition of LERNs (resp. LBRNs) 
%we are able to predict $R^t$ (resp. $R_0$) based on sufficient conditions imposed on the LERNs (resp. LBRNs). That is, we are able to determine  whether or not there is a healthy/endemic equilibrium in the case of the layered networked SIWS model or if there is an outbreak in the case of layered networked SIWR model by just checking if the LBRNs meet the aforementioned sufficient conditions. Fourth, motivated by the decentralized nature of the LBRNs, we propose a LBRNs-based optimization framework with preventive and corrective control actions for budget-constrained and rate-constrained allocation problems. Based on the assumptions imposed in the allocation cost functions, we are able to provide the solution to both problems as convex geometric programs. Furthermore, since  the LBRNs only need local information by definition, we are able to prove that the optimization framework proposed can be solved using standard dual-decomposition--based distributed algorithms. 

In addition, we provide an auxiliary result on the closed-form solution of $R^0$ for homogeneous spreading (see Corollary~\ref{coro:homogeneous_R0}). We also elaborate on the decomposability of our framework versus other approaches for epidemic mitigation based on network-level metrics: spectral abscissa and $R^0$ optimization approached (see Proposition~\ref{prop:decomp}).

\textit{Article Organization:} The remainder of this paper is organized as follows. In Section~\ref{sec:model}, we introduce the layered networked SIWS and SIWR models. The analysis of the equilibrium and transient behavior subject to threshold conditions based on the local and layer-wise metrics are given in Section~\ref{sec:analysis}. %\jose{Section~\ref{sec:analysis} contains the characterization of the dynamical behavior and equilibrium at the network and node level based on the threshold conditions imposed on the network-level and local metrics}. Also, we provide sufficient conditions to leverage the local reproduction numbers to make predictions of the epidemic spreading at the network-level. 
In Section~\ref{sec:optimization}, we develop an LBRNs-based optimization framework and provide the solutions to rate and budget-constrained allocation problems %by transforming the original the original optimization problems by means of a change of variable and proving there is no optimality gap introduced. 
as convex GPs. We illustrate our results in Section~\ref{sec:simulations} and assess the performance of our optimization framework against the spectral abscissa and $R_0$ optimization %as the size of the layered network increases
. Section~\ref{sec:conclusion} concludes this article.  All the proofs are included in the appendices at the end of the manuscript.

\textit{Notation}: %For any positive integer $$, we use $\mathcal{V^P}$ to denote the set $\{1,2,\dots,n\}$.
The $i$th entry of a vector $x$ is denoted $x_i$. We use $\mathbf{0}$ and $\mathbf{1}$ to denote the vectors, whose entries are equal to $0$ and $1$, respectively, and use $I_n$ to denote an $n\times n$ identity matrix. For any vector $x\in\mathbb{R}^n$, we use $\text{diag}(x)$ to denote the $n\times n$ diagonal matrix whose $i$th diagonal entry equals $x_i$. For a square matrix $M$, we use $\text{diag}(M)$ to zero out the off-diagonal elements of $M$ and $[M]_j$ to denote the element in the $j$th diagonal entry. For any two real vectors $a,b\in\mathbb{R}^n$, we write $a>b$ if $a_i\geq b_i$ and $a\neq b$, and $a\gg b$ if $a_i>b_i$ for all $i\in\{1,\dots,n\}$. For a square matrix $M$, we use %$\sigma(M)$ to denote the spectrum of $M$, 
$\rho(M)$ to denote the spectral radius of $M$. %$\alpha(M)$ to denote the largest real part among the eigenvalues of $M$, 
%and $\lambda_{\max}(M)$ %is the eigenvalue with the largest real part.
% = \text{arg}\max\{\text{Re}(\lambda):\lambda \in \sigma(M)\}$. 
For any variable $x\in\mathbb{R}^n_{>0}$, we denote its upper and lower bound with $\overline{x},~\underline{x}$, respectively, where $0<\underline{x}<x<\overline{x}$. Consider a 2-layer network where the set of nodes is partitioned between two sets $\mathcal{C}^1$ (set of nodes in layer~1 with $\lvert\mathcal{C}^1\rvert=n_1$) and $\mathcal{C}^2$ (set of nodes in layer~2 with $\lvert\mathcal{C}^2\rvert=n_2$). %where they interact within and across layers. 
We use $[M]_{ij}^{(\sigma,\delta)}$, with node $i\in\mathcal{C}^{\sigma},~j\in\mathcal{C}^{\delta}$, where $\sigma,\delta\in\{1,2\}$, to denote the $(i,j)$ entry of a square matrix $M\in\mathbb{R}^{(n_1+n_2)\times(n_1+n_2)}$ whose value depends on the layer membership of both nodes. Specifically, the superscript pair $(\sigma,\delta)$ distinguishes between intra-layer $(\sigma=\delta)$ and inter-layer interactions $(\sigma\neq\delta)$, which in the context of layered epidemic modeling, these interactions are governed by different parameters. More precisely, $[M]_{ij}^{(\sigma,\delta)}=M_{i^{\circ}j^{\circ}}$, where
$(i^{\circ},j^{\circ})=(i,j)$ if $i,j\in\mathcal{V}^1$, $(i^{\circ},j^{\circ})=(i,j+n_1)$ if $i\in\mathcal{V}^1,~j\in\mathcal{V}^2$, $(i^{\circ},j^{\circ})=(i+n_1,j)$ if $i\in\mathcal{V}^2,~j\in\mathcal{V}^1$, and $(i^{\circ},j^{\circ})=(i+n_1,j+n_1)$ if $i,j\in\mathcal{V}^2.$
%$i^{\circ}=i+n_1(\sigma~\text{mod}~2)$ and $j^{\circ}=j+n_1(\delta~\text{mod}~2)$. 
The same notation is extended to the case of a state vector $z$ where the first $n_1$ entries belong to the layer~1 and the last $n_2$ entries to layer~2: $z_i^{(1)}=z_{i}$, i.e., $i\in\mathcal{V}^1$, and $z_i^{(2)}=z_{i+n_1}$, i.e., $i\in\mathcal{V}^2$. 

 \section{Background and Problem Formulation} \label{sec:model}
% \subsection{Optimal adaptive testing strategies}
% The SIR model equations read:
% \begin{align}
% \frac{ds(t)}{dt} &= -\beta(t)s(t)i_u(t) \\
% \frac{di_u(t)}{dt} &= \beta(t)s(t)i_u(t) - \gamma i_u(t) - \eta\theta(t)i_u(t)
% \end{align}
% Subject to:
% \begin{equation}
% \min_{\theta(\cdot)\geq 0} \int_{t=0}^{\infty} \theta(t)dt \quad \text{such that: } i_u(t) \leq i_{\max} \; \forall t.
% \end{equation}

% \section{Conclusion}
% The theoretical insights highlight testing as a tool for controlling an epidemic.

In this section, we develop a continuous-time \textit{layered networked epidemic model} in a human population network (which could be defined as either a networked SIS or SIR model) coupled with an infrastructure network. %This model will be hereafter referred to as the \textit{layered networked model} (see Fig.~\ref{fig:multilayer_model}). 
We provide some preliminaries results about the relationship between the networked epidemic spreading and the network-level reproduction numbers $R^0$ and $R(t)$. Lastly, we state the research questions we aim to analyze in the sequel.

\begin{figure}%[h!]
    \centering
    \includegraphics[width=0.55\linewidth]{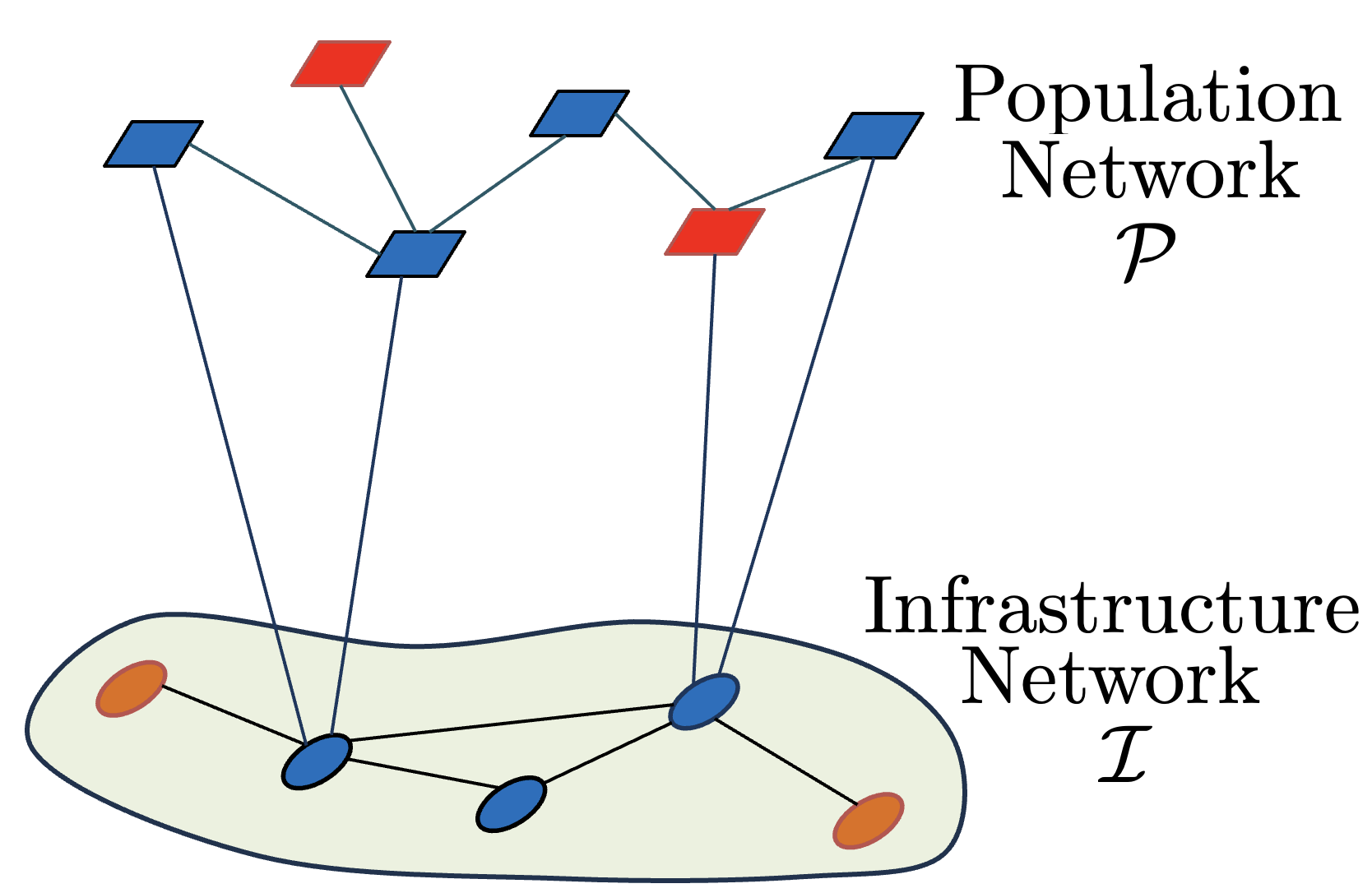}
    \caption{\small Layered networked model with a disease (red) spreading across the population network $\mathcal{P}$ and the pathogen spread %through resource nodes 
    in the infrastructure network $\mathcal{I}$ and any node has two sources of infection, i.e., a group individuals or a resource node.}
    \label{fig:multilayer_model}
\end{figure}

\subsection{Layered Networked Model}
Consider a pathogen spreading over a two-layer network: a population network~$\mathcal{P}$ (layer~1) coupled with an infrastructure network~$\mathcal{I}$ (layer~2). The set of all nodes in the layered network is denoted by $\mathcal{V}$ and it is partitioned into two sets, namely $\mathcal{V^P}$ and $\mathcal{V^I}$. The set of $n$ groups of individuals in the population network is denoted by $\mathcal{V^P}$, i.e., $\lvert \mathcal{V^P}\rvert=n$, and the set of $m$ groups of resources in the infrastructure network is denoted by $\mathcal{V^I}$, i.e., $\lvert \mathcal{V^I}\rvert=m$. We allow any group to be contaminated as a consequence of its interactions with other infected groups of individuals and/or its interactions with infected resource nodes. The contamination $w_j$ at a resource node $j\in\mathcal{V^I}$ evolves as
\begin{align}\label{eq:infra_dyn}
    %\dot{w}_j &= -\gamma_j^ww_j+\sum_{k\in\mathcal{V^I}}\alpha_{kj}w_k - w_j\sum_{k\in\mathcal{V^I}}\alpha_{jk} + \zeta_j^w\sum_{k\in\mathcal{V^P}} c_{jk}^wx_k, \\
    \dot{w}_j &= -\bigg(\gamma_j^w+\sum_{k\in\mathcal{V^I}}\alpha_{kj}\bigg)w_j +\sum_{k\in\mathcal{V^I}}\alpha_{kj}w_k + \zeta_j^w\sum_{k\in\mathcal{V^P}} c_{jk}^wx_k,
\end{align}
where $\gamma_j^w$ denotes the decay rate of contamination at node $j\in\mathcal{V^I}$, $\alpha_{kj}$ represents the flow of the pathogen from node $k$ to node $j$,~for all $k,j\in\mathcal{V^I}$ (with $\alpha_{jj}=0$), $c_{kj}^w$ denotes the  edge weight between a population node $k\in\mathcal{V^P}$ and resource node $j\in\mathcal{V^I}$, and $\zeta_j^w$ denotes the person-to-infrastructure transmission rate at node $j\in\mathcal{V^I}$. %The model in~\eqref{eq:infra_dyn} in vector form becomes
In vector form~\eqref{eq:infra_dyn} becomes
\begin{align}\label{eq:infra_dyn_vector}
    \dot{w} = -\big(\diag(\gamma^w)-A_w\big)w + \diag(\zeta^w)C_wx,
\end{align}
where $\gamma^w=[\gamma_1^w,\dots,\gamma_m^w]$, $A_w$ has negative diagonal entries equal to $-\alpha_j$, with $\alpha_j=\sum_{k\in\mathcal{V^I}}\alpha_{kj}$, and off-diagonal entries equal to $\alpha_{kj}$, and $C_w=[c_{jk}^w]_{m\times n}$. Therefore, the columns of $A_w$ sum to zero, i.e., $A_w\mathbf{1}=\mathbf{0}$ (in other words, there is a conservation of the pathogen flow across each node $j\in\mathcal{V^I}$). Note that the net healing rate of any resource node $i\in\mathcal{V^I}$ is given by the decay rate of contamination $\gamma_j^w$ and the outflow of the contamination $\alpha_j=-[\diag(A_w)]_j$ due to the conservation of the pathogen flow. 
%We consider a networked SIS or networked SIR model with a constant population to study the layered spreading process 

We assume that the layered spreading process evolves in a constant population network
(i.e., %regardless of the compartment model, 
the total number of individuals does not change in time for any node in $\mathcal{V^P}$). To that end, we denote by $s_i,~x_i,~r_i$, the proportion of susceptible, infected, and recovered individuals, respectively, in a group $i\in\mathcal{V^P}$ at time $t\geq 0$. Each group of individuals $i\in\mathcal{V^P}$ has a healing rate $\gamma_i$, a person-to-person infection rate $\beta_i$, and a resource-to-person infection rate $\beta_i^w$. %We use $B\coloneqq[b_{ij}]\in\mathbb{R}_{\geq 0}^{n\times n}$ and $B_w\coloneqq[b_{ij}^w]\in\mathbb{R}_{\geq 0}^{n\times m}$ to denote the adjacency matrix of the population with respect to the population and infrastructure network, respectively. 
For each group of individuals $i\in\mathcal{V^P}$, we use $b_{ij}$ and $b_{ij}^w$ to denote the strength of its incoming connection with another group of individuals $j\in\mathcal{V^P}$ and resource node $j\in\mathcal{V^I}$, respectively. Given the contamination of resource nodes  described by~\eqref{eq:infra_dyn_vector}, the layered networked SIWS model is as follows:
\begin{equation}\label{eq:SIWS}
\begin{aligned}
 \dot{s} &= \diag(\gamma)x - S\big(\diag(\beta)Bx + \diag(\beta^w)B_ww\big), \\
  %\dot{s} &= \big(\diag(\gamma)-S\diag(\beta)B\big)x-S\diag (\beta^w)B_ww, \\
  \dot{x} &= S\big(\diag(\beta)Bx + \diag(\beta^w)B_ww\big) - \diag(\gamma)x,
\end{aligned} 
\tag{SIWS}
\end{equation}
where $\gamma = [\gamma_1,\dots,\gamma_n]$, $\beta=[\beta_1,\dots,\beta_n]$, $S=\diag(s), \beta^w=[\beta_1^w,\dots,\beta_n^w]$, $B=[b_{ij}]_{n\times n}$, and $B^w=[b_{ij}^w]_{n\times m}$ and $s = 1-x$. In the case of a networked SIR model coupled with an infrastructure network with dynamics in~\eqref{eq:infra_dyn_vector}, the resulting layered networked SIWR model is 
\begin{equation}\label{eq:SIWR}
\begin{aligned}
  \dot{s} &= -S\big(\diag(\beta)Bx+\diag(\beta^w)B_ww\big), \\
  \dot{x} &= S\big(\diag(\beta)Bx+\diag(\beta^w)B_ww\big) - \diag(\gamma)x, \\
  \dot{r} &= \diag(\gamma)x,
\end{aligned} 
\tag{SIWR}
\end{equation}
with $s = 1-x-r$. Note that the evolution of the infected state vector $x$ in both \eqref{eq:SIWS} and \eqref{eq:SIWR} depends on the evolution of the susceptible state vector $s$. Consequently, both layered networked models can be written more compactly by introducing
the following notation: 
\begin{align}\label{eq:matrices}
    z&\coloneqq \begin{bmatrix}
        x^{\jose{\top}} & w^{\jose{\top}}
    \end{bmatrix}^{\top},~~~H\big(s\big)\coloneqq \diag\big(\begin{bmatrix}
        s^{\jose{\top}} & \mathbf{1}_m^{\jose{\top}}
    \end{bmatrix}^{\top}\big),%\begin{bmatrix}
        %S(t) & 0 \\ 0 & I_m
    %\end{bmatrix}, 
    \nonumber\\
    B_f &\coloneqq \begin{bmatrix}
        \diag(\beta) B & \diag(\beta^w)B_w \\ \diag(\zeta^w)C_w & A_w - \text{diag}(A_w)
    \end{bmatrix}, \nonumber\\
    D_f &\coloneqq \begin{bmatrix}
        \diag(\gamma) & 0 \\ 0 & \diag(\gamma^w) - \text{diag}(A_w)
    \end{bmatrix}.
\end{align}
Using the notation in~\eqref{eq:matrices}, the dynamics of $z$ for both layered networked models is captured by 
\begin{align}\label{eq:layered_dyn}
    \dot{z} = \big(H(s)B_f-D_f\big)z,
\end{align}
where $s$ is described by either~\eqref{eq:SIWS} or~\eqref{eq:SIWR}. %With a slight abuse of notation, we use $(s,z)^{\text{SIS}}$ to denote the susceptible state vector follows \eqref{eq:SIWS}.  %where $s=1-x$. 
%In the case, the population is modeled with \eqref{eq:SIWR} we use $(s,z)^{\text{SIR}}$ %where $s=1-x-r$
To completely capture the dynamics~\eqref{eq:SIWS} (\eqref{eq:SIWR} resp.) we introduce the layered state vector $(s,z)^{\text{SIS}}$ where $s$ follows~\eqref{eq:SIWS} and~$z$ is given by~\eqref{eq:layered_dyn} ($s$ follows~\eqref{eq:SIWR} resp.).

\subsection{Preliminaries: Network-level Reproduction Numbers}\label{subsec:preliminaries}
\jose{In general,} the basic reproduction number $R^0$ \jose{is the average number of new infections caused by one infectious person in a completely susceptible %proportion
group of individuals. For a standard networked SIS model (without infrastructure coupling), $R^0$ determines whether the system attains an endemic equilibrium or not, i.e., $R^0$-based threshold behavior~\cite{cao}. In the case of the layered networked model~\eqref{eq:SIWS},} it was shown in~\cite{pare2023multilayer} that a threshold behavior similar to that of a standard networked SIS model holds \jose{where $R^0\coloneqq\rho(D_f^{-1}B_f)$, where $D_f$ and $B_f$ are in~\eqref{eq:matrices}. %In the case of 
For a standard networked SIR model, it is known that the epidemic spreading is  always eradicated, i.e., $x^*=0$, and $R^0$ determines whether there is an epidemic outbreak before the spreading is eradicated} or the epidemic immediately dies out~\cite{VANDENDRIESSCHE2017288}.The next~\jose{result elaborates on the equilibrium of the layered networked model~\eqref{eq:SIWR}.} 

\begin{proposition}\label{prop:equilibria_SIWR}
    Suppose Assumption~\ref{assump:parameters}--\ref{assump:coupling} hold. %For the layered networked model in~\eqref{eq:SIWR}, the set of equilibria has the form $(s^*,\mathbf{0})^{\text{SIR}}$, where $s^*\in[0,1]^n$.
    \jose{The set of equilibria for the layered networked model~\eqref{eq:SIWR} has the form $(s^*,\mathbf{0})^{\text{SIR}}\in\mathcal{D}$, where $s^*+r^*=1$.}
\end{proposition}

\begin{comment}
    Proposition~\ref{prop:equilibria_SIWR} implies that the system has an infinite number of healthy equilibria, whereas an endemic equilibria does not exist for \eqref{eq:SIWR} under %given that the healing rates are nonnegative based on
Assumption~\ref{assump:parameters}. Therefore, the threshold behavior that is determined by $R_0$ for \eqref{eq:SIWS}, namely $R_0=\rho(D_f^{-1}B_f)$, does not apply to \eqref{eq:SIWR}. %\jose{In turn, $R_0$ can be used under a slight modification whether there is an epidemic outbreak or the epidemic dies out given any initial condition $(s(0),z(0))^{\text{SIR}}\in\mathcal{D}$.}
\end{comment}

\jose{Proposition~\ref{prop:equilibria_SIWR} implies that the epidemic spreading is always eradicated under~\eqref{eq:SIWR} regardless the initial condition $(s,z)^{\text{SIR}}(0)$, i.e., $z^*=\mathbf{0}$ %as $t\rightarrow\infty$ 
with $\dot{s}$ given by~\eqref{eq:SIWR}. Consequently, each group of individuals $i\in\mathcal{V^P}$ is split between susceptible $s_i^*$ and recovered $r_i^*$ proportions in the equilibrium. %Given that~\eqref{eq:SIWR} does not have an endemic equilibrium, we can not leverage the $R^0$ associated with~\eqref{eq:SIWS} to analyze a possible threshold behavior, i.e., $R^0=\rho(D_f^{-1}B_f)$ is not valid for~\eqref{eq:SIWR}. 
\jose{Intuitively, the layered networked model~\eqref{eq:SIWR} inherits the same properties as the standard networked SIWR epidemic model. Likewise, the basic reproduction number for~\eqref{eq:SIWR} is given by $R^0 \coloneqq\rho(S(0)D_f^{-1}B_f)$ (obtained by standard analysis in networked epidemic processes~\cite{zino,MEI2017116}).}}

\jose{As mentioned in the Introduction, the effective reproduction number $R(t)$ is the network-level metric used to understand the transient growth rate of $z(t)$ in the direction of a vector $v(t)$, where $v(t)$ is the leading eigenvector associated with $\lambda_{\max}(H(s(t))B_f-D_f)$, which determines the direction of growth for the layered infection state. Thus, $R(t)$ is defined as $R(t)\coloneqq \rho(H(s(t))D_f^{-1}B_f)$ capturing the evolution of the susceptible state in the group of individuals (where $s(t)$ either follows~\eqref{eq:SIWS} or~\eqref{eq:SIWR})~\cite{MEI2017116,andrade2022inferring}. In particular, it has been proven that the following threshold behavior exists for $R(t)$ when $s$ follows a SIR epidemic model: when $R(t)>1$, the infection across the layered network is expanding towards its peak, otherwise, the infection is dying out when $R(t)\leq 1$ since some groups of individuals have reached herd immunity.   }

\subsection{Problem Statements}
In the sequel for the model in~\eqref{eq:layered_dyn}, we aim to provide insights related to the following questions:
\begin{enumerate}
    %\item Establish threshold conditions for the equilibria of the layered networked model based on the network-level basic reproduction number $R_0$.
    %\item Define the network-level effective reproduction number $R_0^t$ to characterize the dynamical behavior of a weighted average of $z$.
    \item How can \jose{the local effective reproduction numbers (LERNs)} be defined such that they predict the transient behavior of any node in the layered network according to threshold conditions?
    %\item Identify sufficient conditions to predict the network-level behavior based on assumptions imposed on the LERNs. 
    \item \jose{Under what conditions can the LBRNs be leveraged to predict the network-level epidemic spreading %, i.e., predict the epidemic spread at the network-level,  
    without having to compute $R^0$?}
    % Under what conditions is the network-level spreading process predicted by leveraging the LERNs? Based on LERNs framework, define the local basic reproduction numbers (LBRNs). Do there exist sufficient conditions imposed on the LBRNs to predict the equilibria of the epidemic at a network-level without having to compute $R^0$? 
    %\item Develop two LERNs-based optimal resource allocation problems for the layered network: the \textbf{rate-constrained} allocation that identifies the lowest-cost allocation network-wide to restrict the LERNs below a given upper bound: and the \textbf{budget-constrained} allocation problem, which minimizes the aggregates sum of the LERNs with a limited budget.   
    \item \jose{Assume layer-wise partitions of the layered network: intra- and inter-layer interactions. What is the closed-form solution of the network-level $R^0$ in terms of the physical interactions in the layered network?}
    %\item \jose{Benchmark with other approaches: spectral and network-level basic reproduction number optimization. Not sure given the nature of the layered network.} 
    \item \jose{What is tradeoff in using LBRNs-based GPs for epidemic mitigation versus standard state-of-the-art optimization approacheIs based on network-level metrics?}
\end{enumerate}

\subsection{Positivity Assumptions}
We impose the following assumptions on the parameters of the layered networked system in \eqref{eq:layered_dyn}.
\begin{assumption}\label{assump:parameters}
    For all $i,j \in \mathcal{V^P}$, $\gamma_i>0,~\beta_i>0,~b_{ij}\geq 0$. For all $j,k\in\mathcal{V^I}$, %$\gamma_j^w + \sum_{k\in\mathcal{V^I}}\alpha_{kj}>0$, with
    $\alpha_{kj}\geq 0$, $\gamma_j^w>0$ and $\zeta_j^w>0$. For all $i\in\mathcal{V^P},j\in\mathcal{V^I}$, $b_{ij}^w\geq0$%with at least one pair $(i,j)$ such that $b_{ij}^w>0$
    . For all $j\in\mathcal{V^I},~k\in\mathcal{V^P}$, $c_{jk}^w\geq0$%, with at least one pair $(j,k)$ such that $c_{ij}^w>0$
    . Also, $B$ and $A_w$ are irreducible.
\end{assumption}

Assumption~\ref{assump:parameters} states that the healing and infection rates of each population group $i\in\mathcal{V^P}$ are strictly positive. Also, it ensures that the net decaying of the contamination of any resource node $j\in\mathcal{V^I}$ is strictly positive and results from adding the decaying rate $\gamma_j^w$ plus the outflow of the pathogen $\sum_{k\in\mathcal{V^I}}\alpha_{kj}$ from resource node $j$ (which indeed is equal to its inflow since $A_w\mathbf{1}=0$). Assumption~\ref{assump:parameters} does not guarantee that the population network is coupled with the infrastructure network. Thus, we impose the following assumption. 

\begin{assumption}\label{assump:coupling}
   %There exists 
   At least one population group $i\in\mathcal{V^P}$ and a resource node $j\in\mathcal{V^I}$, such that $b_{ij}^w>0$ and $c_{ij}^w>0$.
\end{assumption}

A natural question to answer is whether the layered networked models above preserve physical interpretability in the context of epidemics modeling. Since $s_i,~x_i,~r_i$ represent proportions of a given group $i\in\mathcal{V^P}$, they must lie in the interval $[0,1]$ for all $t\geq0$. Moreover, the concentration of the virus $w_j$ at each resource node $j\in\mathcal{V^I}$ must remain nonnegative. Otherwise, these states lack physical meaning. Hence, we restrict our analysis to the set 
\begin{align}\label{eq:layered_set}
    \mathcal{D}\coloneqq\{(s,z):s\in[0,1]^n,x\in[0,1]^n,w\in[0,\infty)^m\},
\end{align}
where $(s,z)^{\text{SIS}}\in\mathcal{D}$ with $s=1-x$ and $(s,z)^{\text{SIR}}\in\mathcal{D}$ with $s=1-x-r$. The following result establishes that, under Assumption~\ref{assump:parameters}--\ref{assump:coupling}, the set $\mathcal{D}$ is positively invariant under \eqref{eq:SIWS} and \eqref{eq:SIWR}.
\begin{lemma}\label{lem:positive_invariance}
    Suppose Assumption~\ref{assump:parameters}--\ref{assump:coupling} hold. Suppose that $(s,z)^{\text{SIS}}(0)\in\mathcal{D}$ (respectively, $(s,z)^{\text{SIR}}(0)\in\mathcal{D}$). Then, $(s,z)^{\text{SIS}}(t)\in\mathcal{D}$ (respectively, $(s,z)^{\text{SIR}}(0)\in\mathcal{D}$), for all $t\geq 0$.
\end{lemma} 

% \begin{proof}
%     We provide the sketch proof for \eqref{eq:SIWR} and it can be extended to \eqref{eq:SIWS}. Assume that at some time $\tau$, $s_i,~x_i,~s_i+x_i\in[0,1]$ for all $i\in\mathcal{V^P}$ and $w_j\geq 0$ for all $j\in\mathcal{V^I}$. First consider any node $j\in\mathcal{V^I}$. If $w_j=0$, then from Assumption~\ref{assump:parameters}--\ref{assump:coupling} and \eqref{eq:infra_dyn}, $\dot{w}_j\geq 0$. Thus, $w_j\geq0$ for all $t\geq\tau$. 
    
%     Now consider any node $i\in\mathcal{V^P}$. If at some time $\tau$, $s_i=1,~x_i=r_i=0$, then from~\eqref{eq:SIWR} and Assumptions~\ref{assump:parameters}--\ref{assump:coupling}, $\dot{s}_i\leq0,~\dot{x}_i\geq0$. Therefore, $x_i,r_i\geq0$ for all $t\geq \tau$. %On the other hand, consider 
%     In the case %when 
%     $x_i=1,~s_i=r_i=0$, i.e., node $i\in\mathcal{V^P}$ is totally infected, then from \eqref{eq:SIWR} and Assumptions~\ref{assump:parameters}--\ref{assump:coupling}, $\dot{x}_i<0,~r_i>0$. Therefore, $s_i,r_i\geq0$ for all $t\geq0$. It is straightforward to conclude that when all the individuals in group $i\in\mathcal{V^P}$ are recovered, the susceptible and infected proportions remain at zero, i.e., $s_i=x_i=0,~r_i=1$ for all $t\geq\tau$.

%     Note that the previous arguments hold for any $i\in\mathcal{V^P}$ and any $j\in\mathcal{V^I}$ for $t\geq\tau$. Since by assumption $s_i(0),~x_i(0),~s_i(0)+x_i(0)\in[0,1]$ for all $i\in\mathcal{V^P}$ and $w_j(0)\geq0$ for all $j\in\mathcal{V^I}$, the lemma follows for $\tau=0$. 
% \end{proof}
\section{Local Reproduction Numbers}\label{sec:analysis}

\jose{This section is devoted to establishing local metrics that enable both transient and equilibrium analysis in the layered networked models~\eqref{eq:SIWR} and~\eqref{eq:SIWS}. %namely \textit{local effective reproduction numbers (LERNs)} and \textit{local basic reproduction numbers (LBRNs)}. 
Furthermore, we seek to \jose{identify conditions based} on the proposed local metrics \jose{and the structure of the layered network to characterize the} network-level %threshold behavior analyzed in Section~\ref{subsec:BRN} and \ref{subsec:ERN}. 
reproduction numbers.}

\subsection{Local Effective Reproduction Numbers (LERNs)}

%In order to define a metric that enables the characterization of the spreading process locally (node-level) for the layered networked model \eqref{eq:SIWS} and \eqref{eq:SIWR}, we first 
%\subsubsection{LERNs (local transient behavior)} 
First, we introduce the concept of \textit{effective distributed reproduction numbers} (DRNs). %As in the case of %the network-level ERN
%$R^t$, the concept of DRNs %we are about to introduce 
%applies for both \eqref{eq:SIWS} and \eqref{eq:SIWR}. %Given  the layered networked model in~\eqref{eq:layered_dyn}, we have to define the DRNs for two set of nodes, i.e., the set of group of individuals in the population network $\mathcal{V^P}$ and the set of resource nodes in the infrastructure network $\mathcal{V^I}$.  
In general, the effective DRNs of any node $i\in\mathcal{V}$ is the expected number of secondary infections \jose{due to its interaction with a particular source of contagion (which could be a node either in the population $\mathcal{P}$ or infrastructure $\mathcal{I}$ network).} %Since the DRNs characterize the pairwise interaction between any two nodes $i\in\mathcal{V}^{\sigma},~j\in\mathcal{V}^{\delta}$, where $\sigma,\delta\in\{\mathcal{P},\mathcal{I}\}$, we use the notation $[\cdot]_{ij}^{(\sigma,\delta)}$ to explicitly track the layer membership of each node, allowing the DRNs to be defined %uniformly across all layer combinations 
%without resorting case-by-case interactions in the layered network (see Appendix~\ref{app:0}).} %given the susceptible proportion of node $i\in\mathcal{V}$ at a time instant $t\geq0$. %, where the %the evolution of 
%susceptible proportion at node $i$ is given by $s_i$ at a. %We now define the DRNs for any node $i\in\mathcal{V}$ in the layered networked model. 
\begin{definition}[Effective DRNs]\label{def:DRN_pop}   
    %Let Assumptions \ref{assump:parameters}--\ref{assump:coupling} hold. %$x_i>0$ for all $i\in\mathcal{V^P}$, and $w_j>0$ for all $j \in\mathcal{V^I}$. 
    %\jose{and $s_i>0$ for all $i\in\mathcal{V^P}$.} %For each group of individuals $i\in\mathcal{V^P}$, the DRNs are given by  
    \jose{Given any source of contagion $j\in\mathcal{V}^{\delta}$, the corresponding DRN of a node $i\in\mathcal{V}^{\sigma}$ at a time  $t\geq0$ is expressed as} 
    % \begin{align*}
    %     R_{ij}^t=s_i\frac{\kappa_{ij}}{\gamma_ix_i}~~,~~\kappa_{ij}=\begin{cases}
    %         \beta_ib_{ij}x_j & \text{if}~j\in\mathcal{V^P}\\
    %         \beta_i^wb_{ij}^ww_j & \text{if}~j\in\mathcal{V^I},
    %     \end{cases}
    % \end{align*}
    \jose{\begin{align}\label{eq:DRNs}
        R_{ij}^{(\sigma,\delta)}(t)=\frac{\big[H(s(t))\big]_i^{(\sigma)}\big[B_f\big]_{ij}^{(\sigma,\delta)}}{\big[D_f\big]^{(\sigma)}_i},
    \end{align}}
    %and for each resource node $j \in \mathcal{V^I}$, the DRNs are given by
    % \[ R_{jk}^t= \frac{\kappa^w_{jk}}{(\gamma_j+\alpha_j)w_j}~~,~~
    % \kappa_{jk}^w=\begin{cases}
    %     \zeta_j^wc_{jk}^wx_k &\text{if}~k\in\mathcal{V^P}\\
    %     \alpha_{kj}w_k &\text{if}~k\in\mathcal{V^I},
    % \end{cases}\]
    %where $\alpha_j=- \text{diag}(A_w)_{j}$
    \jose{where $\sigma,\delta\in\{\mathcal{P},\mathcal{I}\}$ and $H(s),~D_f,~ B_f$ are defined in~\eqref{eq:matrices}.}
\end{definition}

 \jose{The DRNs characterize the pairwise interaction between any two nodes accounting for inter and intra-layer interactions. In particular for any population node $i\in\mathcal{V^P}$, the effective DRN also depends on the current proportion of susceptible individuals $s_i$ at time $t\geq 0$ (which could be modeled by~\eqref{eq:SIWR} or~\eqref{eq:SIWS}). %Therefore, each node $i\in\mathcal{V}$ has as many DRNs as neighbors in the two layers has, and 
 Thus, each DRN can be understood as a single population effective reproduction number of node~$i$ with respect to a particular source of contagion. %A very pertinent step after defining the \textit{DRNs} is to a
 By aggregating the interaction of a given node with all of its neighbors in the layered network, %Towards that goal,
%To characterize the spreading process for each node in the layered network model, 
we define the \textit{local effective reproduction numbers (LERNs)} which build off %the DRNs definition. 
Definition~\ref{def:DRN_pop}. However, the ongoing infection at any given node happens at a different scale than the infection of its neighbors. Thus, a normalization is needed with respect to the ongoing infection at a node of interest so the aggregation of all DRNs is physically sound.}
\begin{definition}[LERNs]\label{def:local_ERN}
    %For any node $i\in\mathcal{V}$, the LERN, denoted by $R_i^t$, is the expected number of new infections caused by all possible sources of infection/contamination %given that the proportion of susceptible individuals in node $i$ is evolving with time. More precisely,
    \jose{The LERN of any node $i\in\mathcal{V}^{\sigma},\sigma\in\{\pop,\infra\}$, is the weighted sum of its DRNs given by
    \begin{align}\label{eq:LERNs}
    R_i^{(\sigma)}(t)=\sum_{\delta\in\{\mathcal{P},\mathcal{I}\}}\sum_{j}R_{ij}^{(\sigma,\delta)}(t)\cdot\dfrac{z_j^{(\delta)}}{z_i^{(\sigma)}}.%~~~\sigma\in\{\mathcal{P},\mathcal{I}\}.
    \end{align}}
\end{definition}

% \begin{remark}
%     \jose{Both the DRNs and LERNs are defined as local metrics to characterize the transient behavior at the node level.  Similarly, the intuition behind the structure of the LERNs can be used to define a metric that characterizes the equilibrium at the node-level, namely \textit{local basic reproduction numbers (LBRNs)}.}

%     Note that the unscaled DRNs for the nodes in the population network only depend on the evolution of the susceptible proportion $s_i$. By aggregating all the unscaled DRNs in~\jose{eq} with $s_i=1$, we obtain the \textit{local basic reproduction number (LBRN)} for the population nodes denoted by $R_i^0$ (which is the analogous to $R_0$ at a node-level). %Note that the unscaled DRNs for the resource nodes do not depend on time so 
%     In the same way we compute the LBRNs of the resource nodes using~\jose{eq}.
% \end{remark}

We characterize an LERNs-based threshold behavior for the dynamics of any node $i\in\mathcal{V}$ in the following theorem.
\jose{\begin{theorem}\label{theo:Rt_local}
    Under Assumptions \ref{assump:parameters}--\ref{assump:coupling} and considering an ongoing epidemic spreading, i.e., $z(t)>\mathbf{0}$ at time instant $t\geq0$, the following claims hold for node $i\in\mathcal{V}^{\sigma},~\sigma\in\{\mathcal{P},\mathcal{I}\}:$
    \begin{enumerate}
        \item $R_i^{(\sigma)}(t)>1$ if and only if $z_i^{(\sigma)}$ is increasing.
        \item $R_i^{(\sigma)}(t)<1$ if and only if $z_i^{(\sigma)}$ is decreasing.
    \end{enumerate}
\end{theorem}}
% \begin{theorem}\label{theo:Rt_local}
%     Let Assumption~\ref{assump:parameters} hold. For a given location $i \in \mathcal{V^P}$,
%     % is nonzero, 
%      assume $x_i>0$. For a given resource $j \in \mathcal{V^I}$, assume $w_j>0$. Then, for a given time $t\geq 0$, the following claims hold:
%     \begin{enumerate}
%         \item %For all $i \in \mathcal{V^P}$,
%         $R_i^t>1$ if and only if $x_i$ is increasing, and %For all $j \in \mathcal{V^I}$,
%         $R_j^t>1$ if and only if $w_j$ is increasing. \label{theo:local_ERN_increasing}
%         \item %For all $i\in\mathcal{V^P}$, 
%         $R_i^t<1$ if and only if $x_i$ is decreasing, and %For all $j\in\mathcal{V^I}$, 
%         $R_j^t<1$ if and only if $w_j$ is decreasing. \label{theo:local_ERN_decreasing}
%     \end{enumerate}
% \end{theorem}
% \begin{lemma}
%     Let time $\tau_i\geq0$ satisfy $R_i(\tau_i)<1$ for a location $i\in\mathcal{V^P}$. Then, $R_i(t)<1$, for $t>\tau_i$.
% \end{lemma}

% \begin{corollary}
%     For each location $i\in\mathcal{V^P}$ there exists a unique time $\Tilde{\tau}_i\geq0$ such that $R_i(\Tilde{\tau}_i)=1$. Then, the peak infection for location $i$ occurs at $t=\tilde{\tau}_i$.
% \end{corollary}
\begin{remark}\label{re:conjecture}
    Note that in Theorem~\ref{theo:Rt_local} all the statements hold with strict inequality. That is, we cannot guarantee that when $R_i(t)=1$, a node $i\in\mathcal{V}$ has reached a peak infection or further conclude it is unique. 
\end{remark}

In what remains of this section we aim to leverage the LERNs to infer conclusions about the epidemic spreading at the network-level. More precisely, given some assumptions imposed on the LERNs, we characterize a threshold behavior to predict %the network-level ERN
$R^t$. \jose{Towards that end, we construct the layered effective reproduction matrix by accounting for all possible inter and intra-layer infection contributions in terms of the effective DRNs in~\eqref{eq:DRNs}}. %, i.e., person-to-person, resource-to-person, person-to-resource, resource-to-resource.
%Now, we define the effective reproduction matrix of the multilayer network $\mathcal{R}(t)$ based on \eqref{eq:unscaled_pop} and \eqref{eq:unscaled_infra}. 
%Let the matrix $\mathcal{R}_\mathcal{P}^t$ denote the DRNs of each node $i\in\mathcal{V^P}$ associated with all nodes $j\in\mathcal{V^P}$, i.e., $[\mathcal{R}_\mathcal{P}^t]_{ij}=\tilde{R}_{ij}^t$ for all $i,~j\in\mathcal{V^P}$. In the same way, the matrix $\mathcal{R}^t_\mathcal{I\rightarrow P}$ accounts for the infection contribution from all the resource nodes %in the infrastructure network 
%to the population nodes, i.e., $[\mathcal{R}^t_\mathcal{I\rightarrow P}]_{ij} = \tilde{R}_{ij}^t$ for all $i\in\mathcal{V^P}$,  $j\in\mathcal{V^I}$. Likewise, the matrices $\mathcal{R}^t_\mathcal{P\rightarrow I}$ and $\mathcal{R}^t_\mathcal{I}$ are defined. 
%In the same way, the contamination of the resource nodes from each location in the population network is given by $[\mathcal{R_{P\rightarrow I}}(t)]_{jk}=\Tilde{R}_{jk}(t)$ for all $j\in\mathcal{V^I}$ and $k\in\mathcal{V^P}$. Finally, $[\mathcal{R_I}(t)]_{jk}=\Tilde{R}_{jk}(t)$ for all $j,k\in \mathcal{V^I}$, accounts for the contamination between resource nodes in the infrastructure network. 
%We now introduce the effective reproduction matrix of the layered networked model%, denoted by $\mathcal{R}^t$

\begin{definition}[Layered Effective Reproduction Matrix] \label{def:Rt_matrix}
    \jose{Given an ongoing epidemic spreading, i.e., $z(t)>\mathbf{0}$ at a time instant $t\geq 0$,} the effective reproduction matrix $\mathcal{R}(t)$ of the layered network is given by 
    \[
    \jose{\mathcal{R}(t) = \begin{bmatrix}
        R_{ij}^{(\mathcal{P},\mathcal{P})}(t) & R_{ij}^{(\mathcal{P},\mathcal{I})}(t) \\
        R_{ij}^{(\mathcal{I},\mathcal{P})}(t) & R_{ij}^{(\mathcal{I},\mathcal{I})}(t)
    \end{bmatrix}. }
    \]
    % \[
    % \mathcal{R}^t = \begin{bmatrix}
    %     \mathcal{R}_\mathcal{P}^t & \mathcal{R}^t_\mathcal{I\rightarrow P} \\
    %     \mathcal{R}^t_\mathcal{P\rightarrow I} & \mathcal{R}_\mathcal{I}^t
    % \end{bmatrix}, 
    % \]
    %where $\mathcal{R}_\mathcal{P}^t\in \mathbb{R}^{n\times n}$, $\mathcal{R}^t_\mathcal{I\rightarrow P}  \in \mathbb{R}^{n\times m}$, $\mathcal{R}^t_\mathcal{P\rightarrow I}  \in \mathbb{R}^{m\times n}$ and $ \mathcal{R}_\mathcal{I}^t \in \mathbb{R}^{m\times m}$. 
\end{definition}

% The matrix $\mathcal{R_P}(t)$ accounts for the DRN's of all groups $i\in\mathcal{V^P}$ associated with all groups $j\in\mathcal{V^P}$. In other words, $\mathcal{R_P}(t)$ describes the infection contribution between all the locations in the population network in a distributed fashion. More precisely, the $ij\text{th}$ entry of $\mathcal{R_P}(t)$ is given by $[\mathcal{R_P}(t)]_{ij}=\Tilde{R}_{ij}(t)$ for all $i,j\in\mathcal{V^P}$. On the other hand, the matrix $\mathcal{R_{I\rightarrow P}}(t)$ accounts for the infection contribution from all the resource nodes in the infrastructure network to the population nodes, i.e., $[\mathcal{R_{I\rightarrow P}}(t)]_{ij} = \Tilde{R}_{ij}(t)$ for all $i\in\mathcal{V^P}$ and all $j\in\mathcal{V^I}$. In the same way, the contamination of the resource nodes from each location in the population network is given by $[\mathcal{R_{P\rightarrow I}}(t)]_{jk}=\Tilde{R}_{jk}(t)$ for all $j\in\mathcal{V^I}$ and $k\in\mathcal{V^P}$. Finally, $[\mathcal{R_I}(t)]_{jk}=\Tilde{R}_{jk}(t)$ for all $j,k\in \mathcal{V^I}$, accounts for the contamination between resource nodes in the infrastructure network. 

%By looking at Definition~\ref{def:Rt_matrix}, 
From \eqref{eq:matrices}, %\eqref{eq:unscaled_pop}, and \eqref{eq:unscaled_infra}, 
it is straightforward to conclude that $\mathcal{R}(t)=H(s)D_f^{-1}B_f$. Hence, we can obtain the network effective reproduction number by computing the spectral radius of %the global effective reproduction matrix
$\mathcal{R}(t)$, i.e., $R(t)=\rho(\mathcal{R}(t))$. We leverage the connection between the LERNs and the structure of %the global effective reproduction matrix
$\mathcal{R}(t)$, \jose{to predict $R(t)$ (consequently the transient behavior of the weighted average of $z$)} based on node-level assumptions. To that end, we present the following theorem.

\begin{theorem}\label{theo:node_to_network}
    %Let Assumptions~\ref{assump:parameters}--\ref{assump:coupling} hold and there is \jose{an epidemic spreading in the layered network}  %system has not reached a healthy equilibrium
    \jose{Consider an ongoing epidemic spreading under Assumptions 1--2}, i.e., $z(t)>\mathbf{0}$ for a given time $t\geq0$. Then, for $\sigma\in\{\pop,\infra\}$, following statements hold:
    \jose{\begin{enumerate}
        \item If $\min_{i\in\mathcal{V}^{\sigma}}R_i^{(\sigma)}(t)>1$, then $\rho(\mathcal{R}(t))>1$, \label{theo:all_great_1}
        \item If $\max_{i\in\mathcal{V}^{\sigma}}R_i^{(\sigma)}(t)<1$, then $\rho(\mathcal{R}(t))<1$, \label{theo:all_less_1}
        \item If %$\min_{i\in\mathcal{V}}R_i^{(\sigma)}(t)=\max_{i\in\mathcal{V}}R_i^{(\sigma)}(t)=1$, then $R(t)=1$. \label{theo:all_1}
        $R_i^{(\sigma)}(t)$ is the same for all $i\in\mathcal{V}^{\sigma}$, then $\rho(\mathcal{R}(t))=R_i^{(\sigma)}(t)$. 
    \end{enumerate}}
\end{theorem} 

% \begin{remark}
%     Based on the approach used to construct $\mathcal{R}^t$, we can also build the layered basic reproduction matrix $\mathcal{R}_0$ and further conclude the same threshold behavior for $R_0$ based on assumptions imposed to the LBRNs. That is, based on the definition of the LBRNs we can draw conclusions about the equilibria of~\eqref{eq:SIWS} and the outbreak of~\eqref{eq:SIWR}. 
% \end{remark}

\subsection{Local and Layer-wise Basic Reproduction Numbers} \label{subsubsec:LBRNs}
\jose{Following the same insight used to define the LERNs, we define local metrics, namely \textit{local basic reproduction numbers (LBRNs)}, to predict the network-level basic reproduction number $R^0$. Specifically, the LBRNs can be leveraged to determine whether or not~\eqref{eq:SIWS} attains an endemic equilibrium, or if there is an epidemic outbreak for~\eqref{eq:SIWR}.
\begin{remark}\label{re:LBRNs}
    Since $R^0=\rho(D_f^{-1}B_f)$ for~\eqref{eq:SIWS}  and $R^0=\rho(D_f^{-1}H(s(0))B_f)$ for~\eqref{eq:SIWR}, we obtain the basic DRNs $R_{ij}^{(\sigma,\delta)}$ for~\eqref{eq:SIWS} and ~\eqref{eq:SIWR}, by evaluating $[H(s(t))]_i^{(\sigma)}=I$ in~\eqref{eq:DRNs} and $[H(s(t))]_i^{(\sigma)}=[H(s(0))]_i^{(\sigma)}$, respectively. %$R_{ij}^{(\sigma,\delta)}=[B_f]_{ij}^{(\sigma,\delta)}/[D_f]_i^{(\sigma)}$ for~\eqref{eq:SIWS} and $R_{ij}^{(\sigma,\delta)}=[H(s(0))]_i^{(\sigma)}[B_f]_{ij}^{(\sigma,\delta)}/[D_f]_i^{(\sigma)}$ for~\eqref{eq:SIWR} 
    The physical interpretation of the LBRNs  follows directly from $R^0$, i.e., $R_{ij}^{(\sigma,\delta)}$ denotes the expected number of secondary cases produced by a single infection/contamination in $i\in\mathcal{V}^{\sigma}$ given its interaction with a particular source of infection~$j\in\mathcal{V}^{\delta}$ under a completely susceptible population $s_i=1$ for~\eqref{eq:SIWS} (or accounting the initial condition of $s_i(0)$ for~\eqref{eq:SIWR}.)
\end{remark}} %the effective DRNs %in~\eqref{eq:DRNs} 
\jose{In what follows, we provide necessary and sufficient conditions to charaterize $R^0$ in terms of the LBRNs. 
\begin{proposition}\label{prop:gap_LBRNs}
    Consider~\eqref{eq:SIWS} or~\eqref{eq:SIWR} under Assumptions~\ref{assump:parameters}--\ref{assump:coupling}. Then either one of the following statements hold:
    \begin{enumerate}
        \item $\min_{i\in\mathcal{V}}R_i^{(\sigma)}<R^0<\max_{i\in\mathcal{V}}R_i^{(\sigma)}$, for all $\sigma\in\{\mathcal{P},\mathcal{I}\}$ ~\label{prop:gap1}
        \item $R_i^{(\sigma)}=R^0$ for any $i\in\mathcal{V}^{\sigma},~\sigma\in\{\mathcal{P},\mathcal{I}\}$. ~\label{prop:gap2} %if and only if the LBRNs are the same for all nodes in the layered network.
    \end{enumerate}
\end{proposition}
The main takeaway from Proposition~\ref{prop:gap_LBRNs} is that the network-level basic reproduction number~$R^0$ is lower and upper bounded by the smallest and biggest LBRNs among nodes in all layers. Likewise,~\ref{prop:gap1}) also characterizes the extreme cases when $\min_{i\in\mathcal{V}}R_i^{(\sigma)}>1$ (which implies $R^0>1$) and $\max_{i\in\mathcal{V}}R_i^{(\sigma)} <1$ (which implies $R^0<1$). Statement~\ref{prop:gap2}) states the necessary and sufficient conditions for $R^0$ to be computed by the LBRNs. Intuitively, if all nodes %in the layered network 
have the same LBRN, then $R^0$ can be computed using the local %parameters and in-degree 
information (rates and in-degree) of any node in the layered network (note there is no homogeneity assumption imposed on the parameters of all the nodes). An interesting case arises when %is regularity among nodes with respect to their pair-wise interaction
the aggregation of basic DRNs over %all nodes in 
interactions with either of the layers %$\delta\in\{\mathcal{P},\mathcal{I}\}$ 
is the same for all nodes %$i\in\mathcal{V^{\sigma}}, 
in a given layer $\sigma\in\{\mathcal{P},\mathcal{I}\}$. We categorize the basic DRNs between intra- and inter-layer interactions considering the following assumption. %For the sake of presentation, we denote the aggregation of basic DRNs over intra-layer interactions as $\text{Intra}(\sigma)$ for any $\sigma\in\{\mathcal{P},\mathcal{I}\}$, i.e., $\text{Intra}(\mathcal{P})=\sum_jR_{ij}^{(\mathcal{P},\mathcal{P})}$. Similarly, the aggregation of basic DRNs over inter-layer interactions is denoted by $\text{Inter}(\sigma,\delta)$ for any $\sigma,\delta\in\{\mathcal{P},\mathcal{I}\}$ and $\sigma\neq\delta$.     %, i.e., $\sum_jR_{ij}^{(\mathcal{\sigma},\mathcal{P})}$ and $\sum_jR_{ij}^{(\mathcal{\sigma},\mathcal{I})}$ is the same for all nodes $i\in\mathcal{V}^{\sigma}$ in the same layer $\sigma\in\{\mathcal{P},\mathcal{I}\}$.
\begin{assumption}\label{assump:assump3}
    The basic DRNs for nodes in layers $\sigma,\delta \in \{\mathcal{P},\mathcal{I}\}$ are such that:
    \begin{enumerate}
        \item \textit{Intra-layer ($\sigma=\delta$)}: For each node $i\in\mathcal{V}^{\sigma}$, the aggregation of its basic DRNs %of all nodes in a layer $ \sigma$ 
        over interactions within the same layer $\sigma$ is 
        given by $R_i^{(\sigma,\sigma)}=\sum_jR_{ij}^{(\sigma,\sigma)}$. Furthermore, for all nodes $i\in\mathcal{V}^\sigma$, $R_i^{(\sigma,\sigma)}\eqqcolon\text{Intra}(\sigma)$.
        \item \textit{Inter-layer ($\sigma\neq\delta$)}: For each node $i\in\mathcal{V}^{\sigma}$, the aggregation of its basic DRNs %of all nodes in a layer $\sigma\in\{\mathcal{P},\mathcal{I}\}$ 
        over interactions with a different layer %$\delta\in\{\mathcal{P},\mathcal{I}\}$ $(\sigma\neq\delta)$,
        $\delta$ is given by $R_i^{(\sigma,\delta)}=\sum_jR_{ij}^{(\sigma,\delta)}$. Furthermore, for all nodes $i\in\mathcal{V}^\sigma$, where $\sigma\neq\delta$, $R_i^{(\sigma,\delta)}\eqqcolon\text{Inter}(\sigma,\delta)$.
    \end{enumerate}
\end{assumption}
Based on Assumption~\ref{assump:assump3} we characterize $R^0$ in terms of both types of interactions: inter and intra-layer.
\begin{theorem}\label{theo:layer_pair-wise}
    Let Assumptions~\ref{assump:parameters}--\ref{assump:assump3} hold. %Also, consider that the LBRNs for all nodes $i\in\mathcal{V^{\sigma}}$ in a particular layer $\sigma\in\{\mathcal{P},\mathcal{I}\}$ are such that $\sum_jR_{ij}^{(\sigma,\delta)}$ is the same with respect to layer $\delta\in\{\mathcal{P},\mathcal{I}\}$. 
    Then, the network-level basic reproduction number $R^0$ is given by 
    \begin{align}\label{eq:BRN_closedform}
        R^0 &= \dfrac{\sum_{\sigma}\text{Intra}(\sigma)+\bigg[\Big(\sum_{\sigma}\text{Intra}(\sigma)\Big)^2-4\phi\bigg]^{\frac{1}{2}}}{2}
        %R^0 &=\dfrac{\sum_{\sigma}\text{Infra}(\sigma)+\Bigg[\bigg(\sum_{\sigma}\text{Infra}(\sigma)\bigg)^2-4\bigg(\prod_{\sigma}\text{Infra}(\sigma)-\prod_{\sigma\neq\delta}\text{Inter}(\sigma,\delta)\bigg)\Bigg]^{\frac{1}{2}}}{2}
    \end{align}
    where 
    \[\phi = \prod_{\sigma}\text{Intra}(\sigma)-\prod_{\sigma\neq\delta}\text{Inter}(\sigma,\delta). \]
    %and $\text{Intra}(\sigma)=\sum_jR_{ij}^{(\sigma,\sigma)}$,~$\text{Inter}(\sigma,\delta)=\sum_jR_{ij}^{(\sigma,\delta)}$.
\end{theorem}
The result provided by Theorem~\ref{theo:layer_pair-wise} explicitly highlights the non-trivial relationship between \jose{basic} DRNs of nodes across different layers. In particular, %the closed-form characterization of $R^0$ relies on the fact the pairwise interaction at the layer level, enabling the use of \phil{statement~\ref{prop:gap2}) in Proposition~\ref{prop:gap_LBRNs}} to compute $R^0$.
\jose{since all nodes in a given layer have the same reproduction numbers per layer interaction (Assumption~\ref{assump:assump3}), we leverage Proposition~\ref{prop:gap_LBRNs} to characterize $R^0$.} An immediate result can be obtained when the parameters are homogeneous across nodes in either of the layers.
\begin{corollary}~\label{coro:homogeneous_R0}
    Under Assumptions~~\ref{assump:parameters}--\ref{assump:assump3} and homogeneous transmission and recovery parameters for nodes in each layer, $R^0$ is the spectral radius of the following $2\times2$ matrix
    \[\mathcal{R}^0_h=\begin{bmatrix}
        \dfrac{\beta_h}{\gamma_h}b & \dfrac{\beta^w_h}{\gamma_h}b^w \\ 
        \dfrac{\zeta^w_h}{\hat{\gamma}^w_h}c^w & \dfrac{\alpha_h}{\hat{\gamma}^w_h}
    \end{bmatrix},\]
    where $b=\sum_{j\in\mathcal{V^P}}b_{ij},~b^w=\sum_{j\in\mathcal{V^I}}b_{ij}^w$ for all $i\in\mathcal{V^P}$, and $c^w=\sum_{j\in\mathcal{V^P}}c^w_{ij},~\alpha_h=\sum_{j\in\mathcal{V^I}}\alpha_{ij},~\hat{\gamma}^w_h=\gamma^w_h+\alpha_h$ for all $i\in\mathcal{V^I}$.
\end{corollary}}

 %Observe that the results in Corollary~\ref{coro:LBRN} are central to map the in-degree and the parameters of each node (healing rate, rates of infection/contamination) to infer about the %steady state 
 %equilibrium of the layered networked model. Also, note that the region where the LBRNs have different values around the threshold $R_i^0=1$ for all $i\in\mathcal{V}$ is not explored in Corollary~\ref{coro:LBRN}  (%the same region is not analyzed 
 %resp. for the case of the LERNs in Theorem~\ref{theo:node_to_network}). The gap in Theorem~\ref{theo:node_to_network} and Corollary~\ref{coro:LBRN} is explored through simulations. 

\section{Optimization Framework via LBRNs} \label{sec:optimization}
 Equipped with the definition of the LBRNs and their relationship with the equilibria of the layered networked models: \eqref{eq:SIWS} or \eqref{eq:SIWR}, we now propose a LBRN-based optimization framework for spreading mitigation. All the proofs for the results of the current section are provided in Appendix~\ref{app:2}.
 
 \subsection{LBRNs-Based Epidemic Mitigation}

 In an \jose{analogous manner} to~\cite{preciado2014,preciado2,smith,mai}, we consider two forms of LBRNs-based mitigation problems: \textit{budget-constrained} and \jose{\textit{LBRNs-constrained allocation}. For any node $i\in\mathcal{V}^{\sigma},\sigma\in\{\pop,\infra\}$, we define the decision set $\theta_i^{(\sigma)}\coloneqq\{\text{T}_i^{(\sigma)},\text{H}_i^{(\sigma)}\}$, where $\text{T}_i^{(\sigma)}$ is the set of transmission rates to be reduced (preventive control) and $\text{H}_i^{(\sigma)}$ is the healing rate to be boosted (corrective control). %at node $i\in\mathcal{V}^{\sigma},\sigma\in\{\pop,\infra\}$. 
 For instance, in the case of a group of individuals $i\in\mathcal{V}^{\pop}$, we have that $\text{T}_i^{(\pop)}\in\{\beta_i,\beta_i^w\}$ and $\text{H}_i^{(\pop)}=\{\gamma_i\}$ (similarly, $\text{T}_i^{(\infra)}$ and $\text{H}_i^{(\infra)}$ can be defined following the definition of the LBRNs for a resource node $i\in\mathcal{V^I}$).
 %for a resource node $i\in\mathcal{V}^{\infra}$, we have that $\text{T}_i^{(\infra)}=\{\zeta_i^w\},~\text{H}_i^{(\infra)}=\{\gamma_i^w\}$). 
 Also, the elements of $\theta_i^{(\sigma)}$ must satisfy a set of box constraints given by 
 \[\Theta_i^{(\sigma)}=\Big\{u\in\text{T}_i^{(\sigma)}, u>0~\text{and}~\text{H}_i^{(\sigma)}>0%\theta_i^{(\sigma)}>\mathbf{0}
 ~\lvert ~h\big(\theta_i^{(\sigma)}\big)\leq\mathbf{0}\Big\},\]
 where $h(\theta_i^{(\sigma)})\leq\mathbf{0}$ accounts for the upper and lower bounds of the local rates to be optimized at a given node in the layered network.}
 %In a manner analogous to \cite{preciado2014,smith}, we consider two forms of LBRN-based mitigation problems: \textit{budget-constrained allocation}, and $R_i^0$-\textit{constrained allocation}. To effectively formulate the mitigation problems for any node $i\in\mathcal{V}$, we define the decision vector $\theta_i\coloneqq\{\beta(i),\gamma(i)\}$, where $\beta(i)$ is the set of all infection rates to be reduced (preventive control) and $\gamma(i)$ is the healing rate to be boosted (corrective control) at any node $i\in\mathcal{V}$. For the case of a group of individuals $i\in\mathcal{V^P}$, we have that $\beta(i)=\{\beta_i,~\beta_i^w\}$ and $\gamma(i)=\{\gamma_i\}$ (for a resource node $i\in\mathcal{V^I}$, we have that $\beta(i)=\{\zeta_i^w\},~\gamma(i)=\{\gamma_i^w\}$). Furthermore, the decision vector $\theta_i$ must satisfy some collection of constraints  $h_i(\theta_i)\leq \mathbb{0}$ that denote the feasible interval of $\theta_i$, i.e., $\underline{\beta}(i)\leq\beta(i)\leq\overline{\beta}(i)$,~$\underline{\gamma}(i)\leq\gamma(i)\leq\overline{\gamma}(i)$. 
 % We assume two types of allocation to control the epidemic spreading: \begin{inparaenum}
 %     \item preventive allocation that reduces the infection rate whether it is associated with contagion from nodes within the same layer or nodes from another layer, e.g., $\beta_i,~\beta_i^w$ for $i\in\mathcal{V^P}$ ($\zeta_i^w$ for $i\in\mathcal{V^I}$, resp.) and \item corrective resources that boost the healing rate $\gamma_i$ for $i\in\mathcal{V^P}$ ($\gamma_i^w$ for $i\in\mathcal{V^I}$, resp.). 
 % \end{inparaenum}
 The costs of preventive and corrective control are given by the functions \jose{$f_i(\cdot)$ and $~g_i(\cdot)$}, respectively,
 and they must \jose{follow} a specific structure stated in the following assumption. 

\begin{assumption}~\label{assump:cost_functions}
          The cost functions $f_i$ and $g_i$ are \jose{posynomials with negative exponents}. Further %such that
          :
    \begin{enumerate}
        \item \label{assump:f} The preventive \jose{control} function \jose{$f_i(u)$} is %a posynomial and it 
        decreasing in $u\in\jose{\text{T}_i^{(\sigma)}}$. 
        \item \label{assump:g} The corrective \jose{control} function $g_i$ is %a posynomial and it is
        increasing in $\jose{\text{H}_i^{(\sigma)}}$. Further, 
        \jose{$g_i(\cdot)$ is a posynomial in $\Delta-\text{H}_i^{(\sigma)}$ where $\Delta>\max\{\overline{\text{H}}_i^{(\sigma)}\}_{i\in\mathcal{V}^{\sigma},\sigma\in\{\pop,\infra\}}$.}
    \end{enumerate}
    \end{assumption}
    
    \jose{Assumption~\ref{assump:cost_functions} guarantees that the cost functions for both preventive and corrective control show diminishing returns. In our context %of epidemic mitigation 
    the ``returns" are captured by the LBRNs which represent the epidemic burden each node in the layered network experience during an epidemic spreading. In particular, Assumption~\ref{assump:cost_functions}.\ref{assump:f} guarantees that the lack of investment in preventive control increases the transmission rate. %in the epidemic spreading across the network. 
Assumption~~\ref{assump:cost_functions}.\ref{assump:g} introduces the parameter $\Delta$ which acts as a network-wide physical ceiling %for eradicating the disease 
on the healing capacity across the network
preventing the decision-maker from allocating means to nodes with high in-degree (super-spreaders).}

\jose{The centralized metric we use for the LBRNs-based mitigation problems is the aggregate sum of the LBRNs over the nodes in the set %$\mathcal{V}^{\star}\coloneqq\mathcal{V}^{\pop_{\star}}\cup\mathcal{V}^{\infra_{\star}}$, where
$\pop_{\star}\subset\pop$ and $\infra_{\star}\subset\infra$ which denote the group of individuals and resources nodes of interest for the allocation problems, respectively. Under Assumption~\ref{assump:assump3} we define the following geometric programs:}
%We now define the budget-constrained allocation problem where some budget $c>0$ is %available to be allocated 
%deployed across the layered network %and the decision-maker aims to deploy $c$ 
%to %attain the minimum 
%minimize the aggregated sum of the LBRNs of the nodes of interest. 
\begin{problem}[Budget-Constrained Allocation]\label{prob:budget}
    %Let $\mathcal{V}^{\star}\subseteq\mathcal{V}$ denote the set of nodes to be considered for the allocation problem. Let functions $f_i(\beta(i))$ and $g_i(\gamma(i))$ satisfy Assumption~\ref{assump:cost_functions}. 
    Given a budget $\jose{c_{\max}}>0$: %and $h_i(\theta_i)$, we say that $\{\theta_i^{\star}\}_{i\in\mathcal{V}^{\star}}$ is optimal under a budget-constrained allocation if $\{\theta_i^{\star}\}_{i\in\mathcal{V}^{\star}}$ is a minimizer of 
    \jose{\begin{equation}\label{eq:P1}
        \begin{aligned}
            \min_{\substack{\{\theta_i^{(\sigma)}\}_{i\in\mathcal{V}^{\sigma}}\\\sigma\in\{\pop_{\star}\cup~\infra_{\star}\}}}~&\sum_{i\in\mathcal{V}^{\sigma}} R_i^{(\sigma)}\big(\theta_i^{(\sigma)}\big)\\
        \text{s.t.}~~~~~~& \sum_{i\in\mathcal{V}^{\sigma}}f_i(\text{T}_i^{(\sigma)})+g_i(\Delta-\text{H}_i^{(\sigma)})\leq c_{\max}\\
           %~~~~~~~~~~& h_i(\theta_i^{(\sigma)})\leq \mathbb{0},~~ \forall i\in\mathcal{V}^{\sigma}.
           ~~~~~~~~~~& \theta_i^{(\sigma)}\in \Theta_i^{(\sigma)},~~ \forall i\in\mathcal{V}^{\sigma}.
        \end{aligned} 
        \tag{P1}
    \end{equation}}
\end{problem}
%In the $R_i^0$-constrained allocation problem, we identify the cheapest aggregated cost in both preventive and corrective control given that the aggregated sum of the LBRNs of the nodes of interest is less than a desired threshold $\varepsilon>0$. %$R_i^0\leq\varepsilon_i$ where $\varepsilon_i>0$ is an arbitrary threshold for the LBRNs of the node of interest. 
\begin{problem}[LBRNs-Constrained Allocation]\label{prob:LBRNs}
    %Let $\mathcal{V}^{\star}\subseteq\mathcal{V}$ denote the set of nodes to be considered for the allocation problem. Let functions $f(\beta(i))$ and $g(\gamma(i))$ satisfy Assumption~\ref{assump:cost_functions}. 
    Given the threshold $\varepsilon_{\jose{\max}}>0$: %and $h_i(\theta_i)$, we say that $\{\theta_i^{\star}\}_{i\in\mathcal{V}^{\star}}$ is optimal under a $R_i^0$-constrained allocation if $\{\theta_i^{\star}\}_{i\in\mathcal{V}^{\star}}$ is a minimizer of 
    \jose{\begin{equation}\label{eq:P2}
        \begin{aligned}
            \min_{\substack{\{\theta_i^{(\sigma)}\}_{i\in\mathcal{V}^{\sigma}}\\\sigma\in\{\pop_{\star}\cup~\infra_{\star}\}}}~~&\sum_{i\in\mathcal{V}^{\sigma}} f_i(\text{T}_i^{(\sigma)})+g_i(\Delta-\text{H}_i^{(\sigma)}) \\
        \text{s.t.}~~~~~~& \sum_{i\in\mathcal{V}^{\sigma}} R_i^{(\sigma)}\big(\theta_i^{(\sigma)}\big)\leq \varepsilon_{\max}\\
           ~~~~~~~~~~& \theta_i^{(\sigma)}\in \Theta_i^{(\sigma)},~~ \forall i\in\mathcal{V}^{\sigma}.
        \end{aligned}
        \tag{P2}
    \end{equation}}
\end{problem}
\begin{remark}\label{re:opt}
    \jose{Generally, the optimization problems for epidemic mitigation are based on network-level metrics (spectral abscissa~\cite{preciado2,preciado2014,enyioha} or $R^0$~\cite{smith}) and it is considered that all local parameters in all nodes in the network are decision variables (in more stylized scenarios only the healing rates represent decision variables~\cite{mai}). In contrast, the LBRNs-based epidemic mitigation allows the decision-maker to choose which groups of individuals or resources nodes along with their  corresponding set of local rates (captured by $\theta_i^{(\sigma)}$) are considered for the epidemic mitigation.}
    % The two allocation problems address three possible cases for the epidemic mitigation in the layered networked model: \begin{inparaenum}
    %     \item $\mathcal{V}^{\star}=\mathcal{V}$, i.e., all groups of individuals $i\in\mathcal{V^P}$ and all resource nodes $j\in\mathcal{V^I}$ are considered for the allocation,  
    %     \item $\mathcal{V}^{\star}$ includes all the nodes of either of the layers ($\mathcal{V}^{\star}=\mathcal{V^P}$ or $\mathcal{V}^{\star}=\mathcal{V^I}$),  \item $\mathcal{V}^{\star}$ contains a combination of groups of individuals and resource nodes but not necessarily all nodes of either layer. 
    % \end{inparaenum}
\end{remark}

\subsection{Solution to LBRNs-based Allocation Problems}
 
% Given the geometric programs for the two forms of LBRNs-based mitigation problems, we can conclude that Problems~\ref{prob:budget} and~\ref{prob:LBRNs} can be solved via decomposition methods.
\jose{In what follows, we prove that both Problems~\ref{prob:budget} and \ref{prob:LBRNs} can be transformed to geometric programs (GPs). %by introducing a proper change of variables. 
In particular, the cost function associated with the corrective control $g_i(\Delta-\text{H}_i^{(\sigma)})$ is not a posynomial on the decision variable $\text{H}_i^{(\sigma)}$. Towards that end, we define the variable $z_i^{(\sigma)}\coloneqq \Delta-\text{H}_i^{(\sigma)}$ with domain
\[\mathcal{Z}_i^{(\sigma)}=\Big\{z_i^{(\sigma)}>0~\lvert ~\Delta-\overline{\HR}_i^{(\sigma)}\leq z_i^{(\sigma)}\leq \Delta-\underline{\HR}_i^{(\sigma)}\Big\}.\]
} %Thus, it imposes a careful handling since in its original form it cannot be transformed into a convex function. 
\begin{theorem}\label{theo:opt_budget}
    Suppose Assumption~\ref{assump:cost_functions} hold. %Consider a budget-constrained allocation given in~\eqref{eq:P1}. Then,~\eqref{eq:P1} is solved by the following auxiliary geometric program with decision variables $\{\theta_i,\tilde{\gamma}(i)\}_{i\in\mathcal{V}^{\star}}$:
    % Then, the optimal investment in preventive and corrective controls for a node $i\in\mathcal{V}^{\star}$ is given by $f_i(\beta^{\star}(i))$ and $\tilde{g}_i(\tilde{\gamma}^{\star}(i))$, where $\theta_i^{\star},\tilde{\gamma}^{\star}(i)$ are the optimal solution to $\theta_i$ and $\tilde{\gamma}(i)$ in the following GP:
    \jose{Let $c_{\max}>0,~\varepsilon_{\max}>0$, %and $\widetilde{\text{H}}_i^{(\sigma)}\coloneqq \Delta-\text{H}_i^{(\sigma)}$. 
    and $\theta_i^{(\sigma)}\coloneqq\{\text{T}_i^{(\sigma)},\text{H}_i^{(\sigma)}\}$. The following are true: 
    % \begin{align*}
    %     \min_{\{\theta_i,\tilde{\gamma}(i)\}_{i\in\mathcal{V}^{\star}}}~~~~&\sum_{i\in\mathcal{V}^{\star}} R_i^0(\theta_i)\\
    %     \text{s.t.}~~~~~~& \sum_{i\in\mathcal{V}^{\star}}f_i(\beta(i))+\tilde{g}_i(\tilde{\gamma}(i))\leq c\\
    %          ~~~~~~~~~~& \tilde{h}_i(\tilde{\gamma}(i))\leq \mathbb{0}\\
    %        ~~~~~~~~~~& h_i(\theta_i)\leq \mathbb{0}~~,~~ \forall i\in\mathcal{V}^{\star},
    % \end{align*}
    % where $\tilde{h}_i(\tilde{\gamma}(i))$ is a collection of linear inequality constraints properly defined %to recover the optimal solution of Problem~\ref{prob:budget}, 
    % such that $\gamma^{\star}(i)=1-\tilde{\gamma}^{\star}(i)$, i.e., there is no optimality gap introduced with the change of variable.
    \begin{enumerate}
        \item \eqref{eq:P1} and \eqref{eq:P2} can be solved as  GPs, evaluating $g_i(\cdot)$ in the variable $z_i^{(\sigma)}\in\mathcal{Z}_i^{(\sigma)}$ and adding the coupling constraint $z_i^{(\sigma)}+\HR_i^{(\sigma)}\leq\Delta$.
        \item At the optimum (for either allocation problem), the inequality introduced due to the change of variable holds with strict equality, i.e., $\HR_i^{(\sigma),\star}=\Delta-z_i^{(\sigma),\star}$. 
    \end{enumerate}
    }
\end{theorem}
% \begin{remark}\label{rem:solution_LBRNs}
%     The solution to Problem~\ref{prob:LBRNs} follows the same 
% \end{remark}

% We are able to solve Problem~\ref{prob:LBRNs} in a similar fashion as stated next.
% \begin{theorem}\label{theo:opt_LBRNs}
%     Let Assumption~\ref{assump:cost_functions} holds. Consider a $R_i^0$-constrained allocation given in Problem~\eqref{eq:P2}. Then,~\eqref{eq:P2} is solved by the following auxiliary geometric program with decision variables $\{\theta_i,\tilde{\gamma}(i)\}_{i\in\mathcal{V}^{\star}}$:
%     \begin{align*}
%         \min_{\{\theta_i,\tilde{\gamma}(i)\}_{i\in\mathcal{V}^{\star}}}~~~~& \sum_{i\in\mathcal{V}^{\star}}f_i(\beta(i))+\tilde{g}_i(\tilde{\gamma}(i))\\
%         \text{s.t.}~~~~~~&    \sum_{i\in\mathcal{V}^{\star}} R_i^0(\theta_i)\leq \varepsilon\\
%              ~~~~~~~~~~& \tilde{h}_i(\tilde{\gamma}(i))\leq \mathbb{0}\\
%            ~~~~~~~~~~& h_i(\theta_i)\leq \mathbb{0}~~,~~ \forall i\in\mathcal{V}^{\star},
%     \end{align*}
%     where $\gamma^{\star}(i)=1-\tilde{\gamma}^{\star}(i)$ solves~\eqref{eq:P2} with cost 
%     \[\sum_{i\in\mathcal{V}^{\star}}f_i(\beta^{\star}(i))+\tilde{g}_i(\tilde{\gamma}^{\star}(i)).\]
% \end{theorem}
\jose{Theorem~\ref{theo:opt_budget} provides the steps to guarantee that both the budget and LBRNs-constrained problems are transformed into GPs and thus, they can be solved by standard  convex optimization techniques for GPs~\cite{boydConvexOpt}. By proving that indeed~\eqref{eq:P1} and~\eqref{eq:P2} can be solved as GPs, we seek to understand the physical insight from the LBRNs-based epidemic mitigation framework. It is straightforward to see that based on the network-level metric chosen both allocation problems consist of} %the formulation of Problems~\ref{prob:budget} and~\ref{prob:LBRNs}
%Theorem~\ref{theo:opt_budget} and~\ref{theo:opt_LBRNs}, we can look at Problems~\ref{prob:budget} and~\ref{prob:LBRNs} as 
the aggregation of %$n^{\star}$ 
subproblems %(where $\lvert\mathcal{V}^{\star}\rvert=n^{\star}$) 
coupled via a \textit{complicating constraint} %that involves the allocation decision of all  nodes of interest 
~\cite{Boyd_decomp}. %In the case of Problem~\ref{prob:budget}, the complicating constraint 
The complicating constraint in Problem~\ref{prob:budget} is associated with the allocation of the budget $c_{\jose{\max}}$  across all %the 
nodes of interest. For Problem~\ref{prob:LBRNs}, the complicating constraint is associated with  %values that 
the \jose{LBRNs-based centralized metric (the aggregated sum of LBRNs) %LBRNs of the nodes of interest must take to satisfy 
such that is does not exceed the threshold~$\varepsilon_{\max}$}. Thus, we have the following immediate observation.

\begin{proposition}\label{prop:decomp}
    Problems~\ref{prob:budget} and~\ref{prob:LBRNs} can be solved %through decomposition methods. More precisely, they can be solved 
    through a dual-decomposition based distributed algorithm. 
\end{proposition}
The proof of Proposition~\ref{prop:decomp} follows standard steps in obtaining the dual formulation of convex optimization problems~\cite{Boyd_decomp}. \jose{The convergence of distributed algorithms for epidemic mitigation based on network-level metrics depend on consensus constraints and extensive communication between nodes and their neighbors~\cite{mai,enyioha}. In contrast,   %The main takeaway of Proposition~\ref{prop:decomp} is that 
the LBRNs-based distributed algorithm (obtained from the result in Proposition~\ref{prop:decomp}) relies on the communication between group of individuals/resource nodes with a master decision-maker. Consequently, the communication burden is less expensive against distributed algorithms based on network-level metrics. Moreover, the LBRNs-based distributed algorithm inherits a physical interpretation of how resources are allocated in a real setup between a high-level decision maker (master) and smaller agents (subproblems). The master computes a global variable to guide the local decision process of each subproblem, where their decisions can be computed in parallel.}

\section{Numerical Results}\label{sec:simulations}
For our simulations, we \jose{initially consider a population network with 60 groups of individuals and 20 resource nodes in the infrastructure network. The graph used to represent the layered network is such that Assumption~\ref{assump:parameters}--\ref{assump:assump3} holds, i.e., the network is strongly connected, positive parameters, and homogeneous intra- and inter-layer interactions. %Also, the matrices $B,~B_w,~C_w$, and the off are binary. 
}

%consider a population network of 58 groups of individuals in the population network and 20 resource nodes in the infrastructure network. The matrices $B$, $B_w$ and $A_w$ are all binary, i.e., equal to one whenever nodes $i$ and $j$ are neighbors, for all simulations. Also, for simplicity we assume $C_w=B_w^{\top}$.

\subsection{Transient behavior of layered epidemic processes }
We assume a heterogeneous spreading where the parameters for each node $i\in\mathcal{V}$ are picked uniformly at random from the following intervals: for all \jose{$i\in\mathcal{V^P}$, $\gamma_i\in[0.075,0.2]$, $\beta_i\in[0.15,1]$, $\beta_i^w\in[0.02,0.15]$, for all $j\in\mathcal{V^I}$, $\gamma_j^w\in[0.20,0.40]$, $\zeta_j^w\in[0.24,0.32]$.} The range of $\gamma_i$ and $\gamma_j^w$ were chosen such that the range of recovery times %at both group of individuals and the resource nodes
of any node in the layered network is between 5 and 13 days, while the transmission range was tuned so that $R^0>1$ for \jose{either} layered %SIWR and SIWS 
networked models.

\begin{figure}[t!]
    \centering
    % Scales width to 100% of the column, height adjusts automatically
    \resizebox{0.65\columnwidth}{!}{%
        \subimport{./}{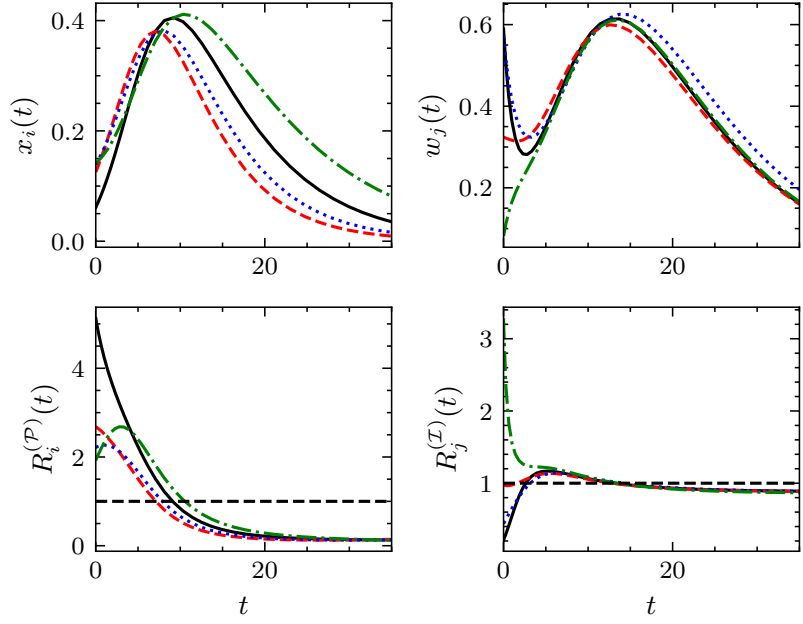}%
    }
    \caption{\small\jose{From left to right: (top) evolution of the infection level $x_i(t),~i\in\mathcal{V^P}$ and $w_j(t),~j\in\mathcal{V^I}$, (bottom) LERNs $R_i^{\pop}(t)$ and  $R_j^{\infra}(t)$ for the groups of individuals and resources nodes considered for the plots at the top. The dash line represents the threshold for the LERNs at one.  Note that the LERNs $R_j^{(\pop)}$ of some resources nodes $j\in\mathcal{V^I}$  cross the threshold more than once.}}
    \label{fig:LERNs}
\end{figure}

\begin{figure}[h]
    \centering
    % Scales width to 100% of the column, height adjusts automatically
    \resizebox{0.65\columnwidth}{!}{%
        \subimport{./}{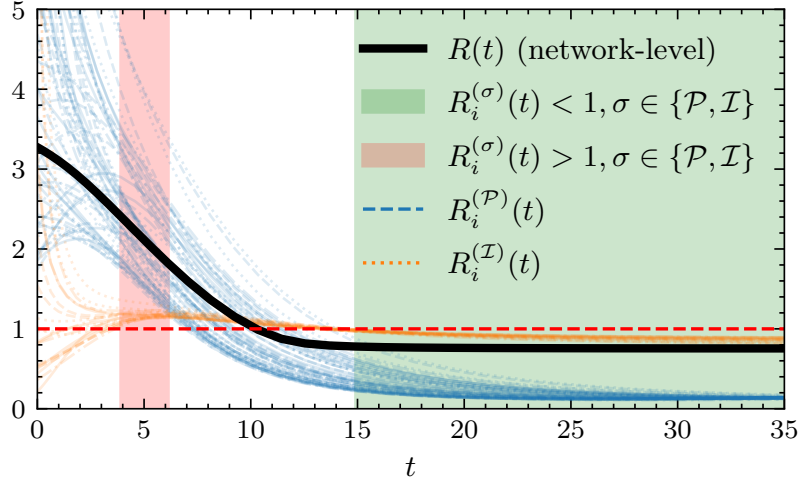}%
    }
    \caption{\small\jose{Network-level effective reproduction number $R(t)$ and the LERNs for all nodes in the layered network. Note that 
    % according to 
    $R(t)$ goes below the threshold (red dashed line at 1) before all of the LERNs are below 1 (green region).}}
    \label{fig:TheoremConference}
\end{figure}

We consider a layered SIWR networked model \jose{to illustrate the relationship between the LERNs and the evolution of the epidemic spreading locally (node-level). The initial condition for $x_i,~i\in\mathcal{V^I}$ and $w_j,~j\in\mathcal{V^I}$ are uniformly picked at random from $[0,0.15]$ and $[0,0.75]$, respectively (we assume $r(0)=\mathbf{0}$)}. Consistent with Theorem~\ref{theo:Rt_local}, %the weighted average $v^{\top}(\tau_p)z(t)$ attains its peak when $R^t=1$ (indicated by the vertical line in the first column of Fig.~\ref{fig:first}). Moreover, $v^{\top}(\tau_p)z(t)$ increases when $R^t>1$ and decreases when $R^t<1$, consistent with %all statements in 
%Theorem~\ref{theo:ERN}. %Consistent with Theorem~\ref{theo:Rt_local}, 
the \jose{infection level} of any node $i\in\mathcal{V^{\sigma},~\sigma\in\{\pop,\infra\}}$, increases when $R_i^{(\sigma)}(t)>1$ and decreases when $R_i^{(\sigma)}(t)<1$, %consistent with Theorem~\ref{theo:Rt_local}
; see Fig.~\ref{fig:LERNs} %(last two columns on the right)
. As noted in Remark~\ref{re:conjecture}, we cannot guarantee that a peak in the transient behavior of any node $i\in\mathcal{V}$ is attained when $R_i^{(\sigma)}(t)=1$. In particular, for some of the resource nodes $j\in\mathcal{V^I}$, it can be seen this is not the case %since $R_i^t=1$ at more than one time instant
\jose{(right bottom plot in Fig.~\ref{fig:LERNs})}. %By inspecting the weighted average $v^{\top}(\tau_p)z(t)$ and the evolution of $w_j$, we can immediately conclude that the dynamical behavior of the resource nodes is 
% totally lost at 
%not captured by
%the network-level metrics. 
%\jose{In Fig.~\ref{fig:second}, we illustrate %the results provided in 
%Theorem~\ref{theo:node_to_network}}.
\jose{Fig.~\ref{fig:TheoremConference} depicts the evolution of the LERNs of all nodes in the layered network (groups of individuals and resource nodes) and the network-level effective reproduction number $R(t)$. } The main insight from Fig.~\ref{fig:TheoremConference} is that leveraging the LERNs to predict the \jose{spreading across} the layered network provides different (arguably better) insight than just \jose{network-level effective reproduction number}, given that $R(t)=1$ much earlier than when all the LERNs are \jose{no longer above one (green region in Fig.~\ref{fig:TheoremConference})}, i.e., the infection is still increasing in some nodes. 

\subsection{Intra- and Inter-layer DRNs: Sensitivity Analysis}

For the sake of illustration, we consider a layered networked SIWS for the rest of the simulations. \jose{In Section~\ref{subsubsec:LBRNs}, we introduce Assumption~\ref{assump:assump3}, which enables the characterization %analysis of the epidemic spreading at the layer level. In particular, Assumption~\ref{assump:assump3} guarantees that the DRNs for the intra and inter-layer interactions is the same for all nodes in a given layer $\sigma\in\{\pop,\infra\}$. Consequently, we are able to characterize 
of the closed-form solution of $R^0$ by reducing the network-level basic reproduction matrix to a 2-by-2 matrix (Theorem~\ref{theo:layer_pair-wise}). We are interested in understanding what is the deviation of  $R^0$ given a layered network does not comply with Assumption~\ref{assump:assump3} from the reproduction number obtained under Assumption~\ref{assump:assump3}. That is, how sensitive is the result in Theorem~\ref{theo:layer_pair-wise}  against some irregularity level $\epsilon\in[0,1]$. Towards that end, we perturb the number of edges of all nodes in the layered network under Assumption~\ref{assump:assump3}. The perturbation depends on: \begin{inparaenum}
    \item a random edge offset $e^{\text{offset}}$ which is uniformly picked at random from an appropriate interval guaranteeing there are no disconnected nodes and that the maximum number of edges a node can have is not exceeded, \item an irregularity level $\epsilon$ which determines the scaling of the sampled edge offset, i.e., $\epsilon=0$ is equivalent to the layered network under Assumption~\ref{assump:assump3} and $\epsilon=1$ means that the sampled edge offset is fully included into the layered network. Therefore, the perturbation is defined as $\lfloor \epsilon e^\text{offset} \rfloor$ (we use the floor operator since the perturbation has to be integers as it represents the number of edges). 
\end{inparaenum} }

\begin{figure}%[htbp]
    \centering
    % Scales width to 100% of the column, height adjusts automatically
    \resizebox{0.65\columnwidth}{!}{%
        \subimport{./}{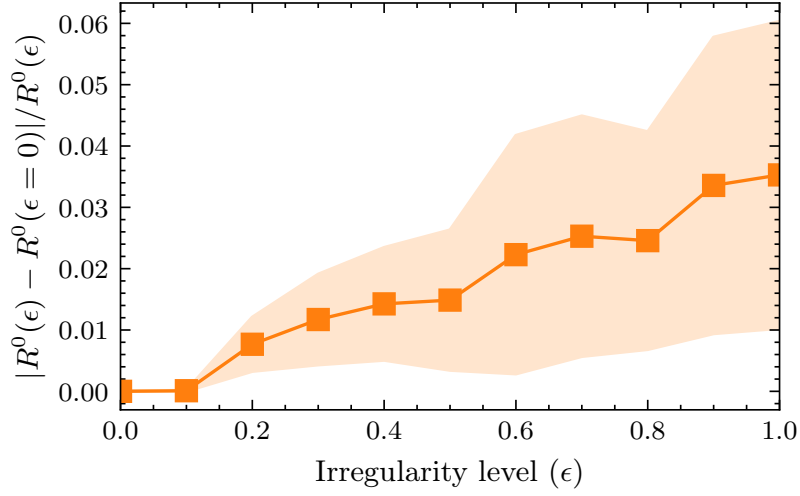}%
    }
    \caption{\small\jose{Relative error of an estimate of $R^0$ under Assumption~\ref{assump:assump3}, where $R^0(\epsilon)$ denotes the network-level basic reproduction number as a function of $\epsilon$. When $\epsilon=0$, the layered network satisfies the conditions in Assumption~\ref{assump:assump3}.}}
    \label{fig:regAnalysis1}
\end{figure}

\jose{Considering the network selected under Assumption~\ref{assump:assump3}, Fig.~\ref{fig:regAnalysis1} depicts the mean (solid line) and the one standard deviation region of the relative error obtained %for the regular network selected  
as the irregularity level $\epsilon$ increases for $100$ realizations of $e^{\text{offset}}$ per value of $\epsilon$. Note that for $\epsilon<0.1$, the relative error is zero which 
% is an effect of the $\lfloor \epsilon e^\text{offset} \rfloor$, i.e., 
happens because 
the scaling of the offset as a function of the floor function is not large enough to perturb the edges of the layered network. As expected, as $\epsilon$ increases, the one standard deviation increases. However, it is noticeable that under $\epsilon=1$ (highest scaling of the edge offset), the mean of the relative error does not surpass the $4\%$. %That is, the approach proposed to compute $R^0$ when the layered network satisfies the conditions imposed on the intra and inter-layer interactions performs relatively well against the perturbation introduced in the edges of the layered network. 
Fig.~\ref{fig:regAnalysis1} illustrates one the main contributions of our work: we can get a relatively good estimate of the network-level $R^0$ by analyzing physical partitions of the layered networked matrix (in our case intra- and inter-layer partitions), introducing a novel way of predicting outbreaks in  networked epidemic processes.}

\subsection{LBRNs-based Epidemic Mitigation and Performance}

\begin{figure}[t!]
    \centering
    % Scales width to 100% of the column, height adjusts automatically
    \resizebox{0.65\columnwidth}{!}{%
        \subimport{./}{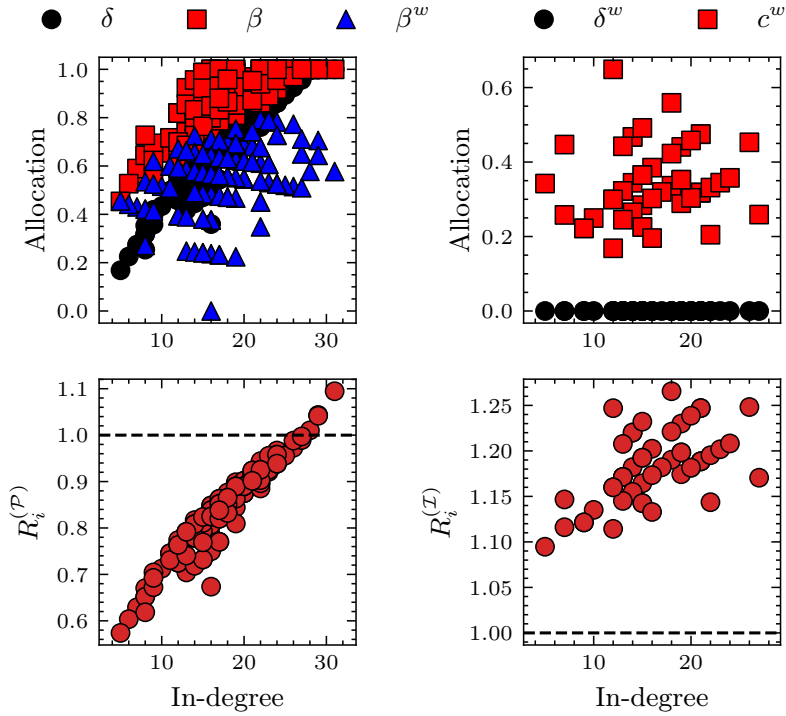}%
    }
    \caption{\small\jose{Budget-constrained allocation (top) for all nodes in the population (left) and the infrastructure (network) with $c_{\max}=330$. (Bottom) LBRNs after the allocation. For the population nodes the allocation is mainly distributed across the healing  $\delta$ and person-to-person transmission rate $\beta$. For the resource nodes, there is no allocation on the healing rate $\delta$ and all the budget is aimed at reducing the person-to-resource contamination rate $c^w$. }}
    \label{fig:allocation2Layers}
\end{figure}

\jose{For the set of simulations in this subsection} we assume a layered networked SIWS model with $150$ groups of individuals and $50$ resources nodes under homogeneous spreading for the sake of results interpretations: \jose{$\gamma_i=0.29$, $\beta_i=0.07$, $\beta_i^w=0.17$, for all $i\in\mathcal{V^P}$, and $\gamma_j^w=0.16$, $\zeta_j^w=0.10$, for all $j\in\mathcal{V^I}$. %Based on this selection of  parameters,
We obtain a pre-intervention %network-level BRN 
$R_0=9.70,~\lambda_{\max}(B_f-D_f)=0.17$ (%the system attains 
there is an endemic equilibrium). \jose{Also,~ $\sum_{i\in\mathcal{V}}R_i^0=1.62\times10^3$, which is the centralized metric for the LBRNs-based epidemic mitigation.  %(resp., for a layered networked SIWR model, %having $R_0>1$ 
%it means there is an outbreak). 
We consider all parameters of all groups of individuals/resource nodes as decision variables, i.e., $\TR_i^{(\pop)}=\{\beta_i,\beta_i^w\},~\HR_i^{(\pop)}=\{\gamma_i\}$, and $\TR_i^{(\infra)}=\{\zeta_i^w\},~\HR_i^{(\infra)}=\{\Delta+\gamma_i^w\}$. We consider the model of cost functions used in~\cite{preciado2014,preciado2,smith}, where} the cost functions %that we use for the allocation of preventive and corrective control actions at any node $i\in\mathcal{V}$ such that they satisfy Assumption~\ref{assump:cost_functions} are given by
$f_i,~g_i$ for any node $i\in\mathcal{V}^{\sigma},~\sigma\in\{\pop,\infra\}$ are given by
\[f_i(\TR_i^{(\sigma)}) = \frac{\big[\TR_i^{(\sigma)}\big]^{-1}-\big[\overline\TR_i^{(\sigma)}\big]^{-1}}{\big[\underline{\TR}_i^{(\sigma)}\big]^{-1}-\big[\overline\TR_i^{(\sigma)}\big]^{-1}},\]
and 
\[g_i(\Delta-\HR_i^{(\sigma)}) = \frac{\big[\Delta-\HR_i^{(\sigma)}\big]^{-1}-\big[\Delta-\underline{\HR}_i^{(\sigma)}\big]^{-1}}{\big[\Delta-\overline{\HR}_i^{(\sigma)}\big]^{-1}-\big[\Delta-\underline{\HR}_i^{(\sigma)}\big]^{-1}},\]}
where %the parameter $\Delta$ adjusts the shape of the cost curve for corrective actions and it is such that 
$\Delta>\max\{\overline{\gamma}(i)\}_{i\in\mathcal{V}}$. Note that both $f_i(\TR_i^{(\sigma)})$ and $g_i(\Delta-\HR_i^{(\sigma)})$ satisfy Assumption~\ref{assump:cost_functions}. \jose{In particular $f_i(\underline{\HR}_i^{\sigma})=1$ accounts for the most aggressive strategies for reducing the transmission rates, like a group of individuals under lockdown or mass testing infrastructure. On the other side, $g_i(\overline{\HR}_i^{(\sigma)})=1$ denotes the most expensive decisions to boost the healing rates, like emergency clinical infrastructure or continuous medical staffing.}

We adopt the perspective in \cite{smith} to choose the intervals of the decision variables based on the pre-intervention parameters: given that %corrective and preventive 
% control actions never decrease the healing rates or increase the transmission rates, we  use $\underline{\gamma}_i=\gamma_i$, $\overline{\beta}_i=\beta_i$, $\overline{\beta}_i^w=\beta_i^w$,  for all $i\in\mathcal{V^P}$, and  $\underline{\gamma}_j^w=\gamma_j^w$, $\overline{\zeta}_j^w=\zeta_j^w$, for all $j\in\mathcal{V^I}$. 
\jose{$\underline{\HR}_i^{(\sigma)}=\HR_i^{(\sigma)},~\overline{\TR}_i^{(\sigma)}=\TR_i^{(\sigma)}$ for all $i\in\mathcal{V}^{(\sigma)},~\sigma\in\{\pop,\infra\}$}. To reflect a 90\% reduction in spreading across the layered network and 50\% reduction in mean recovery time at maximum allocation, we choose %$\overline{\gamma}_i=2\gamma_i$, $\underline{\beta}_i=0.1\beta_i$, $\underline{\beta}_i^w=0.1\beta_i^w$, for all $i\in\mathcal{V^P}$, and $\overline{\gamma}_j^w=2\gamma_j^w$, $\underline{\zeta}_j^w=0.1\zeta_j^w$, for all $j\in\mathcal{V^I}$. 
\jose{$\overline{\HR}_i^{(\sigma)}=2\HR_i^{(\sigma)},~\underline{\TR}_i^{(\sigma)}=0.1\TR_i^{(\sigma)}$ for all $i\in\mathcal{V}^{(\sigma)},~\sigma\in\{\pop,\infra\}$}. Using Theorem~\ref{theo:opt_budget}, we solve a budget-constrained allocation problem with $c_{\max}=330$. \jose{ %the investment on correction and prevention control actions are depicted in Fig.~\ref{fig:allocation2Layers}. 
The results of the LBRNs-based epidemic mitigation are: $R_0=1.10,~\lambda_{\max}(B_f-D_f)=0.17$, $\sum_{i\in\mathcal{V}}R_i^0=1.85\times10^2$. In~Fig.~\ref{fig:allocation2Layers} we depict the allocation of the budget across $\TR_i^{(\sigma)}$ and $\HR_i^{(\sigma)}$ across the nodes in the layered network. The respective optimal LBRNs obtained from the allocation are consistent with Proposition~\ref{prop:gap_LBRNs}: the highest and lowest LBRNs obtained from the optimization %(1.31 and 0.57, respectively) %are 1.311 and 0.57 
define an upper and lower bound for $R^0$: $1.311<R^0<0.57$. Also, note that there is no investment made for the healing rates in the infrastructure network. Consequently, $R_i^{(\infra)}$ are all above the threshold. In contrast, groups of individuals experience  a gradual investment in the healing rates as their in-degree increases, while the majority of the resources has been allocated to the person-to-person transmission rate $\beta_i$.} %The main insights from Fig.~\ref{fig:allocation2Layers} are: \begin{inparaenum}
    %\item the investment in the healing rates of the resource nodes in the infrastructure network (top right plot) is null. Intuitively, the total recovery rate of any node $j\in\mathcal{V^I}$ is given by $\gamma_j^w+\alpha_j$. Thus, the healing rate is already boosted by $\alpha_j$ which is equivalent to the in-flow of contamination into a resource node (per dynamics of $w_j$ give in~\eqref{eq:infra_dyn}). 
    %\item Not necessarily all LBRNs need to be less than 1, to guarantee that the network-level $R_0<1$ (gap in Corollary~\ref{coro:LBRN}). In the example illustrated in Fig.~\ref{fig:LBRNs-example} (bottom right plot), we have that $R_j^0>1$ for all $j\in\mathcal{V^I}$.
%\end{inparaenum}

% \begin{figure}
%     \centering
%     \includegraphics[width=\linewidth]{Extended/Figures_J1/LBRNsAllocationExample.png}
%     \caption{Budget-constrained allocation problem using LBRNs-based optimization. Note that the investment in corrective actions for resource nodes is less predominant than the control actions in the population network.}
%     \label{fig:LBRNs-example}
% \end{figure}

In order to showcase the performance between our optimization framework, %and 
the spectral abscissa~\cite{preciado2,preciado2014}, and $R_0$ minimization~\cite{smith}, we solve %both budget and rate-constrained 
a budget-constrained problem assuming the healing rates of all resources nodes $j\in\mathcal{V^I}$ are fixed. We also increase the %order of the number of nodes in 
size of the layered network to analyze the computational effort of these three approaches. Table~\ref{tab:budget-comparison} summarizes the results obtained for the three aforementioned methods where $\overline{x}^{\star}$ (resp. $\overline{w}^{\star}$) represents the  average of the equilibrium of the infected states in the population network $x^{\star}$ (resp. $w^{\star}$ \jose{in the infrastructure network}) and time is in seconds. As expected, our method outperforms the other two centralized approaches in terms of computational effort, %a feature that becomes more evident as the order of the network \phil{increases} 
and it is a consequence of the decomposability of our approach (consistent with Proposition~\ref{prop:decomp}). \jose{Although our method yields higher $\overline{w}^{\star}$ in the two network instantiations, our method outperforms at least one of the other methods based on network-level metrics. Note that, the $\overline{x}^{\star}$  obtained under our approach and the $R^0$ optimization  is close. Also, the $R^0$ obtained for the spectral abscissa and LRBNs-based optimization do not deviate much. Table~\ref{tab:budget-comparison} showcases the effect of the layered network topology in the results obtained for metric-based algorithm. In turn, the LBRNs-based epidemic mitigation can be situated in the middle of the two centralized methods based on network-level metrics, by just relying on a fundamental notion: the aggregated sum of the burden that each node experiences during an epidemic outbreak, captured by the LBRNs. }  %It is important to note that the performance of these three approaches is sensitive to the connectedness of the layered network. For instance, for the small network $(n=58,~m=20)$, the $R_0$ obtained with our approach ($R_0=1.07$) is higher than the one obtained through the minimization of the spectral abscissa, which is the opposite for the network with $n=580,~m=200$. 
% Another important takeaway is that %the weighted of the equilibrium 
% $\overline{x}^{\star}$ and $\overline{w}^{\star}$ are always higher for our approach, which can be viewed as the optimality lost %in our approach 
% by using the LBRNs compared to the other two centralized approaches. 

\begin{table}[t]
\centering
\footnotesize
\setlength{\tabcolsep}{3.5pt}
\renewcommand{\arraystretch}{1.1}
\begin{tabular}{|c|c|c|c|c|c|c|}
\hline
Network & Method & $\bar{x}^{\star}$ & $\bar{w}^{\star}$ &
$R_0$ & $\lambda_{\max}$ & Time \\
\hline
\multirow{3}{*}{N1}
& LBRNs    & {0.83} & 2.19 & 2.58 & 0.70 & \textbf{0.40} \\
& Abscissa & 0.70 & 0.22  & 2.54 & 0.40 & 1.98 \\
& $R_0$    & 0.87 & 1.00  & 2.02 & 0.66 & 3.10 \\
\hline
\multirow{3}{*}{N2}
& LBRNs    & 0.88 & 8.06 & {1.78} & 0.62 & \textbf{1.044} \\
& Abscissa & 0.77 & 1.41 & 2.44 & 0.23 & 3.54 \\
& $R_0$    & 0.87 & 3.31 & 1.51 & 0.43 & 4.04 \\
\hline
\end{tabular}
\caption{\small\jose{Comparison of budget-constrained allocations for the LBRNs, spectral abscissa, and $R_0$ optimizations with fixed healing rates on the resource nodes. The network N1 has 150 group of individuals and $50$ resources node and budget $c_{\max}=247.5$, N2 has 350 group of individuals and 150 resource nodes with budget $c_{\max}=607.5$.}}
\label{tab:budget-comparison}
\end{table}

Using Theorem~\ref{theo:opt_budget}, we solve an LBRNs-based rate constrained problem with the goal to achieve eradication and compare our proposed method with the other two approaches~\cite{smith,preciado2014}. The desired rates $\varepsilon$ for the abscissa and $R_0$ optimization problems are $-10^{-3}$ and $1$, respectively. For the LBRNs-based optimization, we determine the $\sum_{i\in\mathcal{V}^{\star}}R_i^0$ necessary to guarantee the eradication in an iterative fashion. Table~\ref{tab:rate-comparison} summarizes the results for the eradication of the virus under the three approaches. As expected, under our approach the budget needed to eradicate the %spreading process 
epidemic is higher than under the centralized approaches while the computation time of our method still outperforms the spectral and $R_0$ minimization%approaches
.

\begin{table}[ht]
\centering
\begin{tabular}{|c|c|c|c|c|c|c|}
\hline
\multicolumn{1}{|c|}{Opt.} &
\multicolumn{2}{c|}{N1} &
\multicolumn{2}{c|}{N2} &
\multicolumn{2}{c|}{N3} \\
\cline{2-7}
Method & Budget & Time &
    Budget & Time  & Budget & Time \\
\hline
LBRNs & 159.81 & 0.25 & $1.72~K$ &  3.48  & $14.3~K$  &     8.28 \\
Abscissa & 137.01 & 1.18 & $1.53~K$ & 4.05  & $13.7~K$ & $91*10^2$
            \\
$R_0$ & 135.58 & 1.28 & $1.51~K$ & 4.08  & $13.0~K$ & $92*10^2$  \\
\hline
\end{tabular}
\caption{\small Allocation comparison between LBRNs-based, $R_0$, and spectral abscissa optimization to achieve epidemic eradication for a layered networked SIWS system.}
\label{tab:rate-comparison}
\end{table}

\section{Conclusion}\label{sec:conclusion}
We have analyzed a unifying framework for layered epidemic processes where the spreading in a population network can either be represented as a networked SIS or SIR model and is coupled with an infrastructure network. %with linear dynamics
% In particular, we have provided a comprehensive analysis of the equilibrium and the transient behavior for both the network and node-level by leveraging the standard network-level metrics and the novel LERNs and LBRNs. %We have analyzed the network-wide behavior of the layered networked system by characterizing its equilibrium and predicting its transient behavior based on the network-level effective reproduction number. We have elaborated on the particularities that arise depending on the dynamics of epidemic spreading in the population network. Furthermore, we introduce the novel local-metrics: local effective and basic reproduction numbers (LERNs and LBRNs) that opens a path for the distributed analysis of the transient behavior and equilibria of layered epidemic processes. 
% Furthermore, we provide LERNs and LBRNs-based threshold conditions that enable the prediction of the network-wide behavior.
% Leveraging the concept of LBRNs we propose a distributed optimization framework for epidemic mitigation in the layered networked model we propose. Given the distributed definition of the LBRNs we are able to write LBRNs-based geometric programs for both budget and rate-constrained optimization problems. We have provided numerical experiments to highlight that our proposed method outperforms other centralized approaches, like spectral abscissa or $R_0$ minimization, in terms of computational effort and analyze qualitative differences. 
\jose{Specifically, we have introduced a novel method to understand the epidemic spreading in the layered network: local and layer-level metrics. In particular, we identify sufficient conditions for the LERNs to predict the transient network-level behavior. Furthermore, we introduce the LBRNs to capture the epidemic burden that each node experiences %. As such, the LBRNs are used to 
and provide a bound for $R^0$. %we identify a sufficient condition to compute $R^0$ in terms of the LBRNs. 
Further, we situate the analysis at the middle ground between the node and network-level. Under some conditions imposed on the intra- and inter-layer interactions, we are able to obtain the closed-form solution of $R^0$. Via simulations, we showed that the $R^0$ obtained under the layer-level assumptions does not drastically deviate from the $R^0$ of general networks. Lastly, we develop an LBRNs-based epidemic mitigation framework based on a natural notation: the network-wide spread can be quantified by the aggregation of the LBRNs. Although, our method never outperforms both the spectral abscissa and $R^0$ based optimization approaches simultaneously, it performs second best between these two methods when looking at different metrics and is the fastest of the algorithms.}
One line of future investigation could focus on \jose{looking at different resolutions at which a network is partitioned and the effect on the optimal $R^0$.} %exploiting the infrastructure network and the LERNs for data-driven control where each resource node has partial observability of the infection state of \jose{groups of individuals}. 
Another problem of interest would be %to design incentives such that the resource nodes are used strategically by different hubs in the population network under some capacity constraints.
\jose{to consider robustness to actuation failure and how to devise a correction strategy to guarantee every node %follows the nominal parameters devised by 
\jose{complies with the recommendations obtained from} the LBRNs-based optimization.}

\appendix

\section{Proofs for Section~\ref{sec:analysis}} \label{app:1}

\proofof{Lemma~\ref{lem:positive_invariance}} The proof of Lemma~\ref{lem:positive_invariance} follows standard result of forward invariance based on Nagumo's theorem. A version of the proof analyzing the dynamics at the boundary of $\mathcal{D}$ given in~\eqref{eq:layered_set} under~\eqref{eq:SIWR} was presented in~\cite{caiza2024}. For~\eqref{eq:SIWS}, one can look at~\cite{pare2023multilayer}. \QED

\proofof{Proposition~\ref{prop:equilibria_SIWR}}
The point $(s^*,z^*)^{\text{SIR}}$ is an equilibrium of~\eqref{eq:SIWR} if and only if:
    \begin{align*}
        \mathbf{0} &= -S^*\big(\diag(\beta)Bx^*+B_ww^*\big) \\
        \mathbf{0} &= S^*\big(\diag(\beta)Bx^*+B_ww^*\big) - \diag(\gamma)x^* \\
        \mathbf{0} &= -\big(\diag(\gamma^w)-A_w\big)w^* + \diag(\zeta^w)C_wx^*.
    \end{align*}
    Adding the first \jose{two} equations yields $\mathbf{0}=Dx^*$. Given that $\diag(\gamma)$ is a positive diagonal matrix, then $x^*$ must be $\mathbf{0}$ and $s^*=1-r^*$. Therefore, from the last equation, we obtain %$\big(-\diag(\gamma^w)+A_w\big)w^*=\mathbf{0}$
\begin{align}\label{eq:w_equilibria}
    \big(-\diag(\gamma^w)+A_w\big)w^*=\mathbf{0}.
\end{align}
Note that by the structure of $A_w$, the column sum of $A_w$ equals zero, i.e., $A_w\mathbf{1}=\mathbf{0}$. Therefore, by Assumption~\ref{assump:parameters}, the Gersgorin Disc Theorem gives that all the eigenvalues of $-\diag(\gamma^w)+A_w$ are in the open left half plane and thus, there is no zero eigenvalue. Therefore, the only solution to \eqref{eq:w_equilibria} is $w^*=\mathbf{0}$.
\QED

\proofof{Theorem~\ref{theo:Rt_local}}
Regarding statement \ref{theo:all_great_1}, we will prove the case for an arbitrary node $i\in\mathcal{V^P}$. The same procedure holds for the case of any resource node $j \in \mathcal{V^I}$.
    
    $(\Leftarrow):$ According to Definition~\ref{def:local_ERN}, the effective reproduction number of a location is given by $R_i(t)=\sum_{j\in\mathcal{V}}R_{ij}(t)$%, for all $i \in \mathcal{V^P}$
    . If we evaluate the summation operator, we have that $R_i(t)=\sum_{j\in\mathcal{V^P}}R_{ij}(t) + \sum_{j\in\mathcal{V^I}}R_{ij}(t)$. Based on Definition~\ref{def:DRN_pop}, we can rewrite $R_i(t)$ as 
    \[R_i(t)=\frac{s_i(t)}{\gamma_ix_i(t)}\Bigg(\beta_{i}\sum_{j\in\mathcal{V^P}}b_{ij}x_j(t)+\beta_i^w\sum_{j\in\mathcal{V^I}}b_{ij}^ww_j(t)\Bigg).\]
    Hence, if $R_i(t)>1$, it is true that 
    \begin{align}
        \frac{s_i(t)}{\gamma_ix_i(t)}\Bigg(\beta_i\sum_{j\in\mathcal{V^P}}b_{ij}x_j(t)+\beta_i^w\sum_{j\in\mathcal{V^I}}b_{ij}^ww_j(t)\Bigg) &>1 \label{eq:Ri_pop}\\ 
        s_i(t)\Bigg(\beta_i\sum_{j\in\mathcal{V^P}}b_{ij}x_j(t) + \beta_i^w\sum_{j\in\mathcal{V^I}}b_{ij}^ww_j(t)\Bigg) - \gamma_ix_i(t) &>0, \label{eq:derivative}
    \end{align}
    which implies $\dot{x}_i(t)>0$. Therefore, $x_i(t)$ is increasing. 

    $(\Rightarrow):$ If the infected proportion at node $i$ is increasing, then $\dot{x}_i(t)>0$, which is given by \eqref{eq:derivative}.  By rearranging terms, we obtain \eqref{eq:Ri_pop}. It immediately follows that $R_i(t)>1$, given that $x_i(t)>0$.

    %To prove the decreasing behavior in statement \ref{theo:local_ERN_decreasing}, we follow the same technique as for statement~\ref{theo:local_ERN_increasing}.
\QED

\proofof{Theorem~\ref{theo:node_to_network}}
\jose{To proof the claims in Theorem~\ref{theo:node_to_network}, we use the following result about the spectral radius of irreducible matrices.
\begin{lemma}[\cite{varga}, Theorem~2.2]\label{lem:rhoIrreduc}
    Let $A\geq 0$ be an irreducible $n\times n$ matrix, and let $\mathcal{H}^{\star}$ be the hyperoctant of vectors $x>\mathbf{0}$. Then, for any $x\in\mathcal{H}^{\star}$, either
    \begin{align*}
        &\min_{1\leq i\leq n}~\frac{\sum_{j=1}^n a_{ij}x_j}{x_i} < \rho(A) < \max_{1\leq i\leq n}~\frac{\sum_{j=1}^n a_{ij}x_j}{x_i},\\
        \text{or}&~~\frac{\sum_{j=1}^n a_{ij}x_j}{x_i} = \rho(A)~~\text{for all}~~1\leq i\leq n. \\
    \end{align*}
\end{lemma}}
\jose{ %By construction the layered infection state $z\coloneqq[x~~w]^{\top}$ is nonnegative for all $t\geq0$. 
Theorem~\ref{theo:node_to_network} assumes there is an ongoing epidemic spreading in the layered network for a given $t\geq 0$, i.e., $z\coloneqq[x~~w]^{\top}>\mathbf{0}$. Let $[\mathcal{R}^t]_{ij}^{(\sigma,\delta)}$ denote the $(i,j)$ entry of the layered effective reproduction matrix $\mathcal{R}(t)$, where $\sigma,\delta\in\{\mathcal{P},\mathcal{I}\}$. By construction of the layered network DRNs (Definition~\ref{def:DRN_pop}), the LERNs (Definition~\ref{def:local_ERN}), and Lemma~\ref{lem:rhoIrreduc} it follows that:
\begin{align}\label{eq:bounds_rho}
    \min_{i\in\mathcal{V}}~R_i^{(\sigma)}(t) < \rho(\mathcal{R}(t)) < \max_{i\in\mathcal{V}}~R_i^{(\sigma)}(t)
\end{align}
where the LERNS for each $i\in\mathcal{V}$ is given by:
\[R_i^{(\sigma)}(t)= \dfrac{\sum_{\delta\in\{\mathcal{P},\mathcal{I}\}}\sum_{j} [\mathcal{R}(t)]_{ij}^{(\sigma,\delta)}z_j^{(\delta)}}{z_i^{(\sigma)}}.\]
%%Theorem~\ref{theo:node_to_network} satisfies Lemma~\ref{lem:rhoIrreduc}.
% By Lemma~\ref{lem:rhoIrreduc}, we can immediately conclude the three claims in Theorem~\ref{theo:node_to_network}.
The first two statements in Theorem~\ref{theo:node_to_network} are extremes cases of~\eqref{eq:bounds_rho}. The last statement is a consequence of the LBRNs being the same for all nodes in the layered network.~\QED}

\proofof{Proposition~\ref{prop:gap_LBRNs}}
\jose{By construction, the basic DRNs are obtained by evaluating $z = \mathbf{1}$ in~\eqref{eq:LERNs} and accounting for  Remark~\ref{re:LBRNs}. Thus, we can invoke Lemma~\ref{lem:rhoIrreduc} when $x=\mathbf{1}$. \jose{The two statements in Proposition~\ref{prop:gap_LBRNs} immediately follow.}}

\proofof{Theorem~\ref{theo:layer_pair-wise} and Corollary~\ref{coro:homogeneous_R0}}
\jose{The basic reproduction matrix, denoted by $\mathcal{R}^0$, is computed as follows
\begin{align*}
    \mathcal{R}^0 &= \begin{bmatrix}
        R_{11}^{(\pop,\pop)} & \cdots & R_{1n}^{(\pop,\pop)}  & R_{11}^{(\pop,\infra)} & \cdots &  R_{1m}^{(\pop,\infra)} \\
        \vdots  & \ddots & \vdots  & \vdots & \ddots & \vdots \\
        R_{n1}^{(\pop,\pop)} & \cdots & R_{nn}^{(\pop,\pop)}  & R_{n1}^{(\pop,\infra)} & \cdots &  R_{nm}^{(\pop,\infra)} \\
        R_{11}^{(\infra,\pop)} & \cdots & R_{1n}^{(\infra,\pop)}  & R_{11}^{(\infra,\infra)} & \cdots &  R_{1m}^{(\infra,\infra)} \\
        \vdots  & \ddots & \vdots  & \vdots & \ddots & \vdots \\
        R_{m1}^{(\infra,\pop)} & \cdots & R_{mn}^{(\infra,\pop)}  & R_{m1}^{(\infra,\infra)} & \cdots &  R_{mm}^{(\infra,\infra)}
    \end{bmatrix}.
\end{align*}
By construction, we can conclude that $R^0=\rho(\mathcal{R}^0)$. Under Assumption~\ref{assump:assump3}, we have that 
\[\mathcal{R}^0\mathbf{1}=\begin{bmatrix}
    \big(\text{Intra}(\pop) + \text{Inter}(\pop,\infra)\big)\mathbf{1}_n \\
    \big(\text{Inter}(\infra,\pop) + \text{Intra}(\infra)\big)\mathbf{1}_m
\end{bmatrix}\]
Let $R^{(\pop)}=\text{Intra}(\pop)+\text{Inter}(\pop,\infra)$, i.e., the LBRN of any group of individuals in the population layer $\pop$ and $R^{(\infra)}=\text{Intra}(\infra)+\text{Inter}(\infra,\pop)$, i.e., the LBRN of any resource node in the infrastructure layer $\infra$. By Proposition~\ref{prop:gap_LBRNs}, we have that the layered basic reproduction number $R^0$ is bounded
\[\min\{R^{(\pop)},R^{(\infra)}\}<R^0<\max\{R^{(\pop)},R^{(\infra)}\}.\]
Thus, $R^0$ is the spectral radius of a reduced matrix $\tilde{\mathcal{R}}^0$ defined as follows:
\begin{align*}
    \tilde{\mathcal{R}}^0=\begin{bmatrix}
        \text{Intra}(\pop) & \text{Inter}(\pop,\infra) \\
        \text{Inter}(\infra,\pop) & \text{Intra}(\infra) 
    \end{bmatrix},
\end{align*}
where $R^0$ is the dominant eigenvalue of $\tilde{\mathcal{R}}^0$ and it is obtained by solving the following characteristic equation
\[\lambda^2-\sum_{\sigma\in\{\pop,\infra\}}\intra(\sigma)\lambda + \phi=0,\] where $\phi$ is the determinant of the reduced matrix $\tilde{\mathcal{R}}^0$. Since $\tilde{\mathcal{R}}^0$ is a positive and irreducible matrix by construction, we can immediately conclude  that it has a positive real eigenvalue equal to its spectral radius (Theorem 2.7 in~\cite{varga}). Thus, $\rho(\mathcal{R}^0)$ is the greatest root from the above quadratic equation whose closed-form solution is given by~\eqref{eq:BRN_closedform}.}

\section{Proofs for Section~\ref{sec:optimization}}\label{app:2}
\proofof{Theorem~\ref{theo:opt_budget}}
    \jose{We first consider the case when a decision-maker chooses to optimize both the person-to-resource transmission rate $\zeta_i^w$ and the contamination decay rate $\gamma_i^w$%, i.e., $\theta_i^{(\infra)}=\{\zeta_i^w,\gamma_i^w\}$
    .}
    %We first look at the case of a resource node $i\in\mathcal{V^I}$ with \phil{$\theta_i^{(\sigma)}=\{\zeta_i^w,\gamma_i^w\}$, i.e., the decision-maker aims to optimize both } Since the decision-maker aims to optimize the healing rate $\gamma_i^w$%whereas the denominator of the LBRN of a node $i\in\mathcal{V^I}$ is given by $\gamma_i^w+\alpha_i$
    \jose{Since the total healing rate of any resource node is given by $\text{H}_i^{(\infra)}=\gamma_i^w+\alpha_i$, where $\alpha_i$ is the outflow of the contamination in node $i\in\mathcal{V}^{\infra}$, we let $\theta_i^{(\infra)}=\{\zeta_i^w,\HR_i^{(\infra)}-\alpha_i\}$.} %introduce the following auxiliary variable $\hat{\gamma}_i^w=\gamma_i^w+\alpha_i$. 
%     Thus, %the allocation cost at node $i\in\mathcal{V^I}$ is given by 
%     %\[f_i(\zeta_i^w)+\tilde{g}_i(1+\alpha_i-\hat{\gamma}_i^w)\]
%     %and 
%     the box constraints are given by 
% \phil{\begin{align}\label{eq:const_infra}
%         h_i(\theta_i^{(\infra)})\leq\mathbb{0}=\begin{cases}
%         \underline{\zeta}_i^w\leq \zeta_i^w\leq\overline{\zeta}_i^w\\
%         \underline{\gamma}_i^w\leq {\HR}_i^{(\infra)}-\alpha_i\leq\overline{\gamma}_i^w,~~~i\in\mathcal{V^I}.
%     \end{cases}
%     \end{align}}
    % $\underline{\zeta}_i^w\leq \zeta_i^w\leq\overline{\zeta}_i^w$ and $\underline{\gamma}_i^w\leq \hat{\gamma}_i^w-\alpha_i\leq\overline{\gamma}_i^w$. 
    %So far, this change of variable still does not guarantee that~\eqref{eq:P1} can be transformed into a convex optimization given that $\tilde{g}_i$ is a posynomial but its domain is not strictly a decision variable, i.e., it has $\Delta+\alpha_i$ as an offset. 
    \jose{Following the identity $z_i^{(\infra)}=\Delta+\alpha_i-\HR_i^{(\infra)}$, we have that 
    \[\mathcal{Z}_i^{(\infra)}=\Big\{z_i^{(\infra)}>0~\lvert ~\Delta-\overline{\gamma}_i^w\leq z_i^{(\infra)} \leq \Delta-\underline{\gamma}_i^w\Big\}.\]}%Towards that end, we introduce the auxiliary variable %denoted by 
    %$\tilde{\gamma}_i^w=\Delta+\alpha_i-\hat{\gamma}_i^w$ such that the %allocation 
    %cost at node $i\in\mathcal{V^I}$ is given by $f_i(\zeta_i^w)+\tilde{g}_i(\tilde{\gamma}_i^w)$. Thus, the set of constraints \jose{of any resource} node $i\in\mathcal{V^I}$ in terms of $\tilde{\gamma}_i^w$ are
    % \begin{align*}
    %     \tilde{h}_i(\tilde{\gamma}_i^w)\leq\mathbb{0}\coloneqq\begin{cases}
    %         \hat{\gamma}_i^w+\tilde{\gamma}_i^w\leq \Delta+\alpha_i\\
    %     \Delta-\overline{\gamma}_i^w\leq\tilde{\gamma}_i^w\leq \Delta-\underline{\gamma}_i^w,
    %     \end{cases}
    % \end{align*}
    %where the first constraint is a relaxation of the change of variable. The relaxation is necessary to comply with the structure of a geometric program, i.e., $\sum_if_i(x)\leq 1$, where $f_i(x)$ is a monomial. 
    \jose{Similarly, assuming that for any group of individuals $i\in\mathcal{V}^{\pop}$ the person-to-person, resource-to-person; and the healing rates are decision variables, then $\TR_i^{(\pop)}=\{\beta_i,\beta_i^w\}$ and $\HR_i^{(\pop)}=\{\gamma_i\}$. Also, the identity is $z_i^{(\pop)}=\Delta-\gamma_i$ and the feasible set $\mathcal{Z}_i^{(\pop)}$ is given by:
    \[\mathcal{Z}_i^{(\pop)}=\Big\{z_i^{(\pop)}>0~\lvert ~\Delta-\overline{\gamma}_i\leq z_i^{(\pop)} \leq \Delta-\underline{\gamma}_i\Big\}.\]}
    
    The rest of the proof is oriented to solve~\eqref{eq:P1} (the same arguments can be used on~\eqref{eq:P2}). Thus, the augmented problem with decision variables $\{\theta_i^{(\sigma)},~z_i^{(\sigma)}\}_{i\in\mathcal{V}^{\sigma}},~\sigma\in\{\pop_{\star},\infra_{\star}\}$ is defined as:
    
    \begin{equation}\label{eq:P1_proof}
        \begin{aligned}
            \min_{\substack{\{\theta_i^{(\sigma)},z_i^{(\sigma)}\}_{i\in\mathcal{V}^{\sigma}}\\\sigma\in\{\pop_{\star}\cup~\infra_{\star}\}}}~&\sum_{i\in\mathcal{V}^{\sigma}} R_i^{(\sigma)}\big(\theta_i^{(\sigma)}\big)\\
        \text{s.t.}~~~~~~& \sum_{i\in\mathcal{V}^{\sigma}}f_i(\text{T}_i^{(\sigma)})+g_i(z_i^{(\sigma)})\leq c_{\max},\\
           %~~~~~~~~~~& h_i(\theta_i^{(\sigma)})\leq \mathbb{0},~~ \forall i\in\mathcal{V}^{\sigma}.
           ~~~~~~~~~~& z_i^{(\pop)}+\gamma_i\leq \Delta,\\
           ~~~~~~~~~~& z_i^{(\pop)} \in \mathcal{Z}_i^{(\pop)},~~\forall i\in\mathcal{V}^{\pop_{\star}},\\
           ~~~~~~~~~~& z_i^{(\infra)}+\HR_i^{(\infra)}\leq \Delta +\alpha_i,\\
           ~~~~~~~~~~& z_i^{(\infra)} \in \mathcal{Z}_i^{(\infra)},~~\forall i\in\mathcal{V}^{\infra_{\star}},\\
           ~~~~~~~~~~& \theta_i^{(\sigma)}\in \Theta_i^{(\sigma)},~~ \forall i\in\mathcal{V}^{\sigma}.
        \end{aligned}
        \tag{AUX-\ref{eq:P1}}
    \end{equation}
    So far,~\eqref{eq:P1_proof} falls within the standard structure of a GP in terms of all its decision variables. Note that the solution for $\TR_i^{(\sigma)}$ is the same for the original~\eqref{eq:P1} and the transformed~\eqref{eq:P1_proof} optimization problems since there is no change of variable introduced for the transmission rates. Conversely, the inequality constraints $z_i^{(\pop)}+\gamma_i\leq \Delta$,~$z_i^{(\infra)}+\HR_i^{(\infra)}\leq \Delta +\alpha_i$ are a relaxation of the identities used to define the domain of the cost function $g_i(\cdot)$. It remains to be shown that both inequality constraints hold with strict equality at the optimum.
    % We need to guarantee that at optimality the relaxations %introduced by the change of variables 
    % hold with equality, i.e., there is no optimality gap: $\tilde{\gamma}_i^{\star}=\Delta-\gamma_i^{\star}$ and $\tilde{\gamma}_i^{w,\star}=\Delta+\alpha_i-\hat{\gamma}_i^{w,\star}$. We %will 
    % provide the proof for the case of a group of individuals $i\in\mathcal{V^P}$ and it can be %easily 
    % extended for any resource node $j\in\mathcal{V^I}$. Assume by contradiction $\tilde{\gamma}_i^{\star}+\gamma_i^{\star}<\Delta$. Suppose that $\gamma_i^{\star}=\overline{\gamma}_i$. Thus, by the hypothesis we have $\tilde{\gamma}_i^{\star}<\Delta-\overline{\gamma}_i$ which violates the box constraint of $\tilde{\gamma}_i$ in~\eqref{eq:box2}. Thus, at optimality $\tilde{\gamma}_i^{\star}=\Delta-\gamma_i^{\star}$ when $\gamma_i^{\star}=\overline{\gamma}_i$. The same steps can be followed when $\tilde{\gamma}_i^{\star}=\Delta-\underline{\gamma}_i$. Now, assume $\underline{\gamma}_i<\gamma_i^{\star}<\overline{\gamma}_i$. By hypothesis, we have that $\tilde{g}_i(\tilde{\gamma}_i^{\star})<\tilde{g}_i(\Delta-\gamma_i^{\star})$%(see Fig.~\ref{fig:proof_contra})
    % . By Assumption~\ref{assump:cost_functions}, we can conclude that $\tilde{\gamma}_i^{\star}+\gamma_i^{\star}>1$, which is clearly a contradiction. This concludes the proof.    
    
    \jose{%Since the inequality constraints $z_i^{(\pop)}+\gamma_i\leq \Delta$,~$z_i^{(\infra)}+\HR_i^{(\infra)}\leq \Delta +\alpha_i$ are decoupled 
    First, we prove by contradiction that the constraint $z_i^{(\pop)}+\gamma_i\leq \Delta$ for a node $i\in\mathcal{V}^{\pop}$ holds with  equality. Let $\theta_i^{(\sigma),\star},z_i^{(\sigma),\star},~i\in\mathcal{V}^{\sigma},~\sigma\in\{\pop_{\star},\infra_{\star}\}$ be the optimal solution to~\eqref{eq:P1_proof}. Suppose there is a group of individuals $k\in\mathcal{V}^{\pop}$ such that $z_k^{(\pop),\star}+\gamma_k^{\star}< \Delta~\Rightarrow \gamma_k^{\star}< \Delta-z_k^{(\pop),\star}$. Assume there is a candidate solution for the same group of individuals $\tilde{\gamma}_k,~z_k^{(\pop),\star},$  such that $\tilde{\gamma}_k=\Delta-z_k^{(\pop),\star}$. Since $z_k^{(\pop),\star}\in \mathcal{Z}_k^{(\pop)}$, thus $\tilde{\gamma}_k\in\Theta_k^{(\pop)}$, i.e., the new candidate solution is feasible in~\eqref{eq:P1}. Also, note that the budget constraint in~\eqref{eq:P1_proof} depends strictly on $z_i^{(\sigma),\star}$ and $\TR_i^{(\sigma),\star}$, which are the same for the optimal and candidate solution. Thus, the budget constraint remains satisfied. Recall that the LBRN  is a node-level metric that captures the local epidemic burden and by definition (Remark~\ref{re:LBRNs}) is a strictly decreasing function with respect to the healing rate, i.e., its first derivative is strictly negative. Since $\tilde{\gamma}_k>\gamma_k^{\star}$, the LBRNs for the group of individuals $k$ drops, i.e., $R_k^{(\pop)}(\tilde{\theta}_k^{(\pop)})<R_k^{(\pop)}(\tilde{\theta}_k^{(\pop),\star})$, which is clearly a contradiction. 
    The same logic can be used for any node in the infrastructure network $\infra$. This concludes the proof. \QED}

\bibliographystyle{unsrt} % Options: plain, IEEEtran, unsrt, etc.
\bibliography{refs}

@article{andrade2022inferring,
  title={Inferring the effective reproductive number from deterministic and semi-deterministic compartmental models using incidence and mobility data},
  author={Andrade, Jair and Duggan, Jim},
  journal={PLoS Computational Biology},
  volume={18},
  number={6},
  pages={e1010206},
  year={2022},
  publisher={Public Library of Science San Francisco, CA USA}
}

@book{boydConvexOpt,
  title={Convex Optimization},
  author={Boy, Stephen and Vandenberghe, L.},
  volume={1},
  year={2004},
  publisher={Cambridge University Press}
}

@INPROCEEDINGS{Cianfanelli,
  author={Cianfanelli, Leonardo and Parise, Francesca and Acemoglu, Daron and Como, Giacomo and Ozdaglar, Asuman},
  booktitle={Proc.\ Conference on Decision and Control (CDC)}, 
  title={Lockdown interventions in {SIR} models: Is the reproduction number the right control variable?}, 
  year={2021},
  volume={},
  number={},
  pages={4254-4259},
  doi={10.1109/CDC45484.2021.9682977}}

@ARTICLE{zino,
  author={Zino, Lorenzo and Cao, Ming},
  journal={IEEE Circuits and Systems Magazine}, 
  title={Analysis, Prediction, and Control of Epidemics: A Survey from Scalar to Dynamic Network Models}, 
  year={2021},
  volume={21},
  number={4},
  pages={4-23},
  doi={10.1109/MCAS.2021.3118100}}

@misc{WHO,
  author = {{World Health Organization}},
  title = {The true death toll of {COVID-19}. Estimating global excess mortality},
  year = 2021,
  howpublished = {\url{https://www.who.int/data/stories/the-true-death-toll-of-covid-19-estimating-global-excess-mortality}},
  note = {}
}

@misc{harvard,
  author = {{Harvard Kennedy School}},
  title = {How {COVID-19} has changed public policy},
  year = 2021,
  howpublished = {\url{https://www.hks.harvard.edu/faculty-research/policy-topics/public-leadership-management/how-covid-19-has-changed-public-policy}},
  note = {}
}

@ARTICLE{policyCOVID19,
  author={Croo, Deserai A. and DeLeo, Rob A. and Albright, Elizabeth A. and Taylor, Kristin and Birkland, Tom and Zhang, Manli and Koebele, Elizabeth and Jeschke, Nathan and Shanahan, Elizabeth A. and Cage, Caleb},
  journal={Review of Policy Research}, 
  title={Policy learning and change during crisis: {COVID-19} policy responses across six states}, 
  year={2022},
  volume={40},
  number={1},
  pages={10-35},
  doi={https://doi.org/10.1111/ropr.12511}}

@ARTICLE{pare2023multilayer,
  author={Paré, Philip E. and Janson, Axel and Gracy, Sebin and Liu, Ji and Sandberg, Henrik and Johansson, Karl H.},
  journal={IEEE Trans. on Control of Network Systems}, 
  title={Multilayer {SIS} Model With an Infrastructure Network}, 
  year={2023},
  volume={10},
  number={1},
  pages={295-307},
  doi={10.1109/TCNS.2022.3203352}}

@INPROCEEDINGS{baike2023DRNs,
  author={She, Baike and Paré, Philip E. and Hale, Matthew},
  booktitle={Proc. American Control Conference (ACC)}, 
  title={Distributed Reproduction Numbers of Networked Epidemics}, 
  year={2023},
  volume={},
  number={},
  pages={4302-4307},
  doi={10.23919/ACC55779.2023.10156093}}

@ARTICLE{preciado2014,
  author={Preciado, Victor M. and Zargham, Michael and Enyioha, Chinwendu and Jadbabaie, Ali and Pappas, George J.},
  journal={IEEE Trans. on Control of Network Systems}, 
  title={Optimal Resource Allocation for Network Protection Against Spreading Processes}, 
  year={2014},
  volume={1},
  number={1},
  pages={99-108},
  doi={10.1109/TCNS.2014.2310911}}

@misc{Boyd_decomp,
  author = {Boyd, Stephen and Xiao, Lin and Mutapcic, Almir and Mattingley, Jacob},
  title = {Notes on Decomposition Methods},
  year = {2015},
  url = {https://stanford.edu/class/ee364b/lectures/decomposition_notes.pdf},
  urldate = {2015-05-11} 
}

@INPROCEEDINGS{caiza2024,
  author={Caiza, José I. and Qin, Junjie and Paré, Philip E.},
  booktitle={Proc. Conference on Decision and Control (CDC)}, 
  title={Global and Distributed Reproduction Numbers of a Multilayer {SIR} Model with an Infrastructure Network}, 
  year={2024},
  volume={},
  number={},
  pages={5298-5303},
  doi={10.1109/CDC56724.2024.10886419}}

@ARTICLE{preciado2,
  author={Nowzari, Cameron and Preciado, Victor M. and Pappas, George J.},
  journal={IEEE Trans. on Control of Network Systems}, 
  title={Optimal Resource Allocation for Control of Networked Epidemic Models}, 
  year={2017},
  volume={4},
  number={2},
  pages={159-169},
  doi={10.1109/TCNS.2015.2482221}}

@ARTICLE{smith,
  author={Smith, Kevin D. and Bullo, Francesco},
  journal={IEEE Trans. on Automatic Control}, 
  title={Convex Optimization of the Basic Reproduction Number}, 
  year={2023},
  volume={68},
  number={7},
  pages={4398-4404},
  doi={10.1109/TAC.2022.3212012}}

@ARTICLE{mai,
  author={Mai, Van Sy and Battou, Abdella and Mills, Kevin},
  journal={IEEE Control Systems Letters}, 
  title={Distributed Algorithm for Suppressing Epidemic Spread in Networks}, 
  year={2018},
  volume={2},
  number={3},
  pages={555-560},
  doi={10.1109/LCSYS.2018.2844118}}

@INPROCEEDINGS{enyioha,
  author={Enyioha, Chinwendu and Jadbabaie, Ali and Preciado, Victor and Pappas, George},
  booktitle={Proc. European Control Conference (ECC)}, 
  title={Distributed resource allocation for control of spreading processes}, 
  year={2015},
  volume={},
  number={},
  pages={2216-2221},
  doi={10.1109/ECC.2015.7330868}}

@ARTICLE{rowthorn,
  author={Rowthorn, R. E. and Laxminarayan, R. and Gilligan, C. A.},
  journal={Journal of the Royal Society Interface}, 
  title={Optimal control of epidemics in metapopulations}, 
  year={2009},
  volume={6},
  number={41},
  pages={1135-1144},
  doi={10.1109/MCAS.2021.3118100}}

@article{LEE2010136,
title = {Optimal control for pandemic influenza: The role of limited antiviral treatment and isolation},
journal = {Journal of Theoretical Biology},
volume = {265},
number = {2},
pages = {136-150},
year = {2010},
issn = {0022-5193},
author = {Sunmi Lee and Gerardo Chowell and Carlos Castillo-Chávez}
}

@article{HAYHOE2021495,
title = {Multitask learning and nonlinear optimal control of the {COVID-19} outbreak: A geometric programming approach},
journal = {Annual Reviews in Control},
volume = {52},
pages = {495-507},
year = {2021},
issn = {1367-5788},
author = {Mikhail Hayhoe and Francisco Barreras and Victor M. Preciado}
}

@article{ACEMOGLU2024111391,
title = {Optimal adaptive testing for epidemic control: Combining molecular and serology tests},
journal = {Automatica},
volume = {160},
pages = {111391},
year = {2024},
issn = {0005-1098},
author = {Daron Acemoglu and Alireza Fallah and Andrea Giometto and Daniel Huttenlocher and Asuman Ozdaglar and Francesca Parise and Sarath Pattathil}
}

@ARTICLE{somers,
  author={Somers, Vera L. J. and Manchester, Ian R.},
  journal={IEEE Control Systems Letters}, 
  title={Sparse Resource Allocation for Control of Spreading Processes via Convex Optimization}, 
  year={2021},
  volume={5},
  number={2},
  pages={547-552},
  doi={10.1109/LCSYS.2020.3003401}}

@ARTICLE{ashish,
  author={Hota, Ashish R. and Godbole, Jaydeep and Paré, Philip E.},
  journal={IEEE Trans. on Network Science and Engineering}, 
  title={A Closed-Loop Framework for Inference, Prediction, and Control of {SIR} Epidemics on Networks}, 
  year={2021},
  volume={8},
  number={3},
  pages={2262-2278},
  doi={10.1109/TNSE.2021.3085866}}

@INPROCEEDINGS{brooks,
  author={Vrabac, Damir and Shang, Mingfeng and Butler, Brooks and Pham, Joseph and Stern, Raphael and Paré, Philip E.},
  booktitle={Proc. American Control Conference (ACC)}, 
  title={Capturing the Effects of Transportation on the Spread of {COVID-19} with a Multi-Networked {SEIR} Model}, 
  year={2021},
  volume={},
  number={},
  pages={3152-3157}}

@article{PARE2020345,
title = {Modeling, Estimation, and Analysis of Epidemics over Networks: An Overview},
journal = {Annual Reviews in Control},
volume = {50},
pages = {345-360},
year = {2020},
author = {Philip E. Paré and Carolyn L. Beck and Tamer Başar}}

@article{RANDAZZO2020115942,
title = {{SARS}-{C}o{V}-2 {RNA} in Wastewater Anticipated {COVID-19} Occurrence in a Low Prevalence Area},
journal = {Water Research},
volume = {181},
pages = {115942},
year = {2020},
author = {Walter Randazzo and Pilar Truchado and Enric Cuevas-Ferrando and Pedro Simón and Ana Allende and Gloria Sánchez}}

@article{VERMEULEN2015109,
title = {Advancing Waterborne Pathogen Modelling: Lessons from Global Nutrient Export Models},
journal = {Current Opinion in Environmental Sustainability},
volume = {14},
pages = {109-120},
year = {2015},
author = {Lucie C Vermeulen and Nynke Hofstra and Carolien Kroeze and Gertjan Medema}}

@article{CICCOLINI2013380,
title = {Infection prevention in a connected world: The case for a regional approach},
journal = {International Journal of Medical Microbiology},
volume = {303},
number = {6},
pages = {380-387},
year = {2013},
note = {Special Issue Antibiotic Resistance},
author = {Mariano Ciccolini and Tjibbe Donker and Robin Köck and Martin Mielke and Ron Hendrix and Annette Jurke and Janette Rahamat-Langendoen and Karsten Becker and Hubert G.M. Niesters and Hajo Grundmann and Alexander W. Friedrich}}

@ARTICLE{Tien2010-ch,
  title    = "Multiple Transmission Pathways and Disease Dynamics in a
              Waterborne Pathogen Model",
  author   = "Tien, Joseph H and Earn, David J D",
  journal  = "Bulletin of Mathematical Biology",
  volume   =  72,
  number   =  6,
  pages    = "1506--1533",
  month    =  aug,
  year     =  2010
}

@article{SHUAI2011118,
title = {Global Dynamics of Cholera Models with Differential Infectivity},
journal = {Mathematical Biosciences},
volume = {234},
number = {2},
pages = {118-126},
year = {2011},
author = {Zhisheng Shuai and P. {van den Driessche}}}

@INPROCEEDINGS{liu,
  author={Liu, Ji and Paré, Philip E. and Du, Erhu and Sun, Zhiyong},
  booktitle={Proc. American Control Conference (ACC)}, 
  title={A Networked {SIS} Disease Dynamics Model with a Waterborne Pathogen}, 
  year={2019},
  volume={},
  number={},
  pages={2735-2740}}

@INPROCEEDINGS{baike_peak_infection,
  author={She, Baike and Leung, Humphrey C. H. and Sundaram, Shreyas and Paré, Philip E.},
  booktitle={Proc. Conference on Decision and Control (CDC)},
  title={Peak Infection Time for a Networked {SIR} Epidemic with Opinion Dynamics}, 
  year={2021},
  volume={},
  number={},
  pages={2104-2109}}

@article{MEI2017116,
title = {On the dynamics of deterministic epidemic propagation over networks},
journal = {Annual Reviews in Control},
volume = {44},
pages = {116-128},
year = {2017},
author = {Wenjun Mei and Shadi Mohagheghi and Sandro Zampieri and Francesco Bullo}}

@article{VANDENDRIESSCHE2017288,
title = {Reproduction Numbers of Infectious Disease Models},
journal = {Infectious Disease Modelling},
volume = {2},
number = {3},
pages = {288-303},
year = {2017},
author = {Pauline {van den Driessche}}}

@ARTICLE{cao,
  author={Zino, Lorenzo and Cao, Ming},
  journal={IEEE Circuits and Systems Magazine}, 
  title={Analysis, Prediction, and Control of Epidemics: A Survey from Scalar to Dynamic Network Models}, 
  year={2021},
  volume={21},
  number={4},
  pages={4-23}}

@inbook{varga,
    author = {Richard S. Varga},
    title = {Matrix Iterative Analysis},
    publisher = {Springer Berlin, Heidelberg} ,
    year = {2000},
    chapter = {2}
}

@incollection{hethcote1989three,
  title={Three basic epidemiological models},
  author={Hethcote, Herbert W},
  booktitle={Applied Mathematical Ecology},
  pages={119--144},
  year={1989},
  publisher={Springer}
}

@inproceedings{pare2020analysis,
  title={Analysis, online estimation, and validation of a competing virus model},
  author={Par{\'e}, Philip E and Vrabac, Damir and Sandberg, Henrik and Johansson, Karl H},
  booktitle={Proc. American Control Conference (ACC)},
  pages={2556--2561},
  year={2020},
  organization={IEEE}
}

\end{document}